\documentclass[lettersize,journal]{IEEEtran}
\usepackage{amsmath,amssymb,amsfonts}
\usepackage{algorithmic}
\usepackage{algorithm}
\usepackage{array}
\usepackage[caption=false,font=normalsize,labelfont=sf,textfont=sf]{subfig}
\usepackage{textcomp}
\usepackage{stfloats}
\usepackage{url}
\usepackage{verbatim}
\usepackage{graphicx}
\usepackage{cite}
\usepackage{xcolor}
\usepackage{color, colortbl}
\usepackage{multirow}
\usepackage{wasysym}
\usepackage{enumitem}
\usepackage{pifont}
\usepackage{comment}
\usepackage[hidelinks]{hyperref}
\usepackage{adjustbox}
\usepackage{balance}

\newcommand{\eg}{\textit{e.g.,~}}

\newcommand{\ie}{\textit{i.e.,~}}
\newcommand{\etal}{\textit{et al.}}

\newcommand{\defin}[1]{\textcolor{teal}{#1}}

\definecolor{bShade}{RGB}{202, 222, 240}
\definecolor{gShade}{RGB}{151, 189, 226}
\definecolor{rShade}{RGB}{59, 121, 186}

\newcolumntype{B}{>{\columncolor{bShade}}c}
\newcolumntype{G}{>{\columncolor{gShade}}c}
\newcolumntype{R}{>{\columncolor{rShade}}c}

\newcommand{\SECensor}{\textit{Resilient-to-DNS-payload-based-Censorship}}
\newcommand{\ServerCensor}{\textit{Resilient-to-DNS-Server-Censorship}}
\newcommand{\SResReplay}{\textit{Resilient-to-Resolver-Replay-Attack}}
\newcommand{\SANSReplay}{\textit{Resilient-to-ANS-Replay-Attack}}
\newcommand{\SResImp}{\textit{Resilient-to-False-Resolver-Response}}
\newcommand{\SANSImp}{\textit{Resilient-to-False-ANS-Response}}
\newcommand{\SKeyExp}{\textit{Avoid-Duplicating-Long-term-Secret}}

\newcommand{\AResDoS}{\textit{Resilient-to-Resolver-DoS}}
\newcommand{\AANSDoS}{\textit{Resilient-to-ANS-DoS}}

\newcommand{\PConceal}{\textit{Conceal-DNS-Msg-Nature-in-s1}}

\newcommand{\PDecouple}{\textit{Hides-Client-IPaddr-From-Resolver}}
\newcommand{\PHidesTwo}{\textit{Hides-Client-IPaddr-in-s2}}
\newcommand{\PEavesOne}{\textit{Resilient-to-Eavesdropping-in-s1}}
\newcommand{\PEavesTwo}{\textit{Resilient-to-Eavesdropping-in-s2}}

\usepackage{etoolbox}
\makeatletter
\patchcmd{\@makecaption}
  {\scshape}
  {}
  {}
  {}
\makeatother

\def\BibTeX{{\rm B\kern-.05em{\sc i\kern-.025em b}\kern-.08em
    T\kern-.1667em\lower.7ex\hbox{E}\kern-.125emX}}

\definecolor{lightgray}{gray}{0.9}

\newcommand{\headrow}[1]{\multicolumn{1}{c}{\adjustbox{angle=70,lap=\width-0.5em}{#1}}}

\begin{document}

\title{A Survey and Evaluation Framework for Secure DNS Resolution}

\author{
    \IEEEauthorblockN{Ali Sadeghi Jahromi, AbdelRahman Abdou, and Paul C. van Oorschot\\Carleton University, Ottawa, Canada}
}

\maketitle

\begin{abstract}
  Since security was not among the original design goals of the Domain Name System (herein called Vanilla DNS), many secure DNS schemes have been proposed to enhance the security and privacy of the DNS resolution process. Some proposed schemes aim to replace the existing DNS infrastructure entirely, but none have succeeded in doing so. In parallel, numerous schemes focus on improving DNS security without modifying its fundamental two-stage structure. These efforts highlight the feasibility of addressing DNS security as two distinct but compatible stages. We survey DNS resolution process attacks and threats and develop a comprehensive threat model and attack taxonomy for their systematic categorization. This analysis results in the formulation of 14 desirable security, privacy, and availability properties to mitigate the identified threats. Using these properties, we develop an objective evaluation framework and apply it to comparatively analyze 12 secure DNS schemes surveyed in this work that aim to augment the properties of the DNS resolution process. Our evaluation reveals that no single scheme provides ideal protection across the entire resolution path. Instead, the schemes tend to address a subset of properties specific to individual stages. Since these schemes targeting different stages of DNS resolution are complementary and can operate together, combining compatible schemes offers a practical and effective approach to achieving comprehensive security in the DNS resolution process.
\end{abstract}


\section{Introduction}
\label{sec:Intro}
\IEEEPARstart{I}{n} the 1980s, the Domain Name System (DNS) was designed as a distributed, hierarchical name database for Internet-connected resources to replace the \texttt{Hosts.txt} files, which were used to keep track of hostnames and their corresponding IP addresses~\cite{rfc1034, rfc1035}. DNS is a critical component of the Internet; almost all interactions over the Internet begin with a DNS resolution to obtain the IP address associated with a domain name. As the Internet expanded, many entities, such as individuals, IoT devices, and applications, began to rely on domain names to communicate with their intended services. Like other foundational Internet protocols (\eg HTTP), DNS was designed and implemented without security~\cite{rfc1034, rfc1035}. Consequently, for over three decades, the security weaknesses of DNS have been exploited by different classes of adversaries. Governments block or manipulate DNS messages to serve political agendas. For example, censorship in countries such as China is enforced at the Internet Service Provider (ISP) level, or through intermediate devices that manipulate DNS responses to redirect clients to arbitrary destinations or unresponsive addresses~\cite{aryan2013internet, censorIndia, niaki2020iclab, I2Pcensor, censorPakistan, dnsmanipulation}. \texttt{MORECOWBELL} and \texttt{QUANTUMDNS} were projects deployed by the NSA to monitor web and DNS traffic around the world, and to hijack DNS queries~\cite{morecow2017nsa}. As a result, large-scale DNS surveillance and manipulation have significantly undermined the privacy and security of DNS clients worldwide.

Building other protocols and ecosystems on top of DNS amplifies security concerns. For example, the proliferation of IoT devices that rely on DNS has created an ecosystem vulnerable to DNS-related weaknesses. It is estimated that more than 41 billion IoT devices will be connected to the Internet by 2030~\cite{40IoT}. These devices often resolve domain names to communicate with their cloud backends. The lack of privacy in DNS enables intermediate entities, such as ISPs, Autonomous Systems (ASes), and recursive resolvers, to identify the types of IoT devices that a user owns~\cite{apthorpe2017closing}. Furthermore, the absence of robust security mechanisms in DNS makes it vulnerable to active attacks, including pharming~\cite{pharmingGuide}, Kaminsky’s DNS cache poisoning~\cite{KaminskyAttack}, DNS data manipulation~\cite{dnsmanipulation}, reflection and amplification attacks~\cite{ampattack2013dns}, and the exploitation of DNS weaknesses to compromise other protocols and ecosystems~\cite{DNSonTor, jeitner2020impact, hoang2020k, dai2021ip}. Such attacks can redirect IoT devices to malicious cloud or update servers. In addition, botnets, such as Mirai~\cite{understandingmirai}, have abused DNS infrastructure for large-scale reflection and amplification attacks~\cite{AkamaiDNSmirai}.

To secure DNS and mitigate existing threats, various solutions with different approaches have been proposed, including DNSSEC~\cite{rfc4033}, DNS-over-TLS (DoT)~\cite{DoT-rfc7858}, and POPS~\cite{afek2025pops}. These schemes typically protect specific stages of the DNS resolution process, leaving other threats unresolved in different parts of the system. Some secure-DNS approaches also depend on particular infrastructures, such as the web PKI in DoT or the root trust anchor and chain of trust in DNSSEC. These dependencies introduce additional deployability challenges, operational overhead, and potential vulnerabilities inherent to the underlying infrastructure. Secure-DNS schemes often involve trade-offs between the level of security and privacy they provide and the added complexity or performance overhead. An example is DNS-over-Tor (DoTor), which offers strong security and privacy guarantees, but incurs noticeable latency and computational cost. Designing a secure and practical DNS scheme therefore requires carefully balancing the security benefits against the additional complexity and overhead to ensure that the protections are proportionate and the system remains deployable and usable. Achieving this balance increases the likelihood of adoption by relevant stakeholders.

The partial absence of security in the DNS resolution path results in privacy leaks and security weaknesses, and there is no comprehensive overview of the protections that existing secure DNS schemes provide. Privacy concerns remain a key challenge, as companies such as Mozilla, Comcast, and Google have promoted secure recursive resolvers to strengthen client privacy. However, not all recursive resolvers remove privacy-sensitive information, and the use of schemes such as DoT or DNS-over-HTTPS (DoH) does not prevent centralized resolvers from accessing client-related data. This access allows them to construct detailed models of user browsing behavior for purposes such as targeted advertising or client identification based on resolution patterns. Security-conscious clients may wish to prevent such leakage, as DNS queries can reveal sensitive personal information to multiple entities in the resolution process. Despite the importance of these concerns, we are not aware of any evaluation framework that systematically captures the security, privacy, and availability benefits and limitations of secure DNS schemes. An objective and comprehensive evaluation framework is needed to clarify the protections offered by these schemes, facilitate direct comparison, and clearly identify trade-offs to support informed adoption and deployment decisions.

As an initial step toward addressing the security and privacy challenges in the DNS resolution process, there is a need for a comprehensive threat model that systematically identifies the existing threats in the name resolution path. To address this gap, we analyze the weaknesses of the name resolution process by formulating a comprehensive threat model and an attack taxonomy. This model provides a foundation for identifying the primary threats in the DNS resolution process. Building on these findings, we define a set of security, privacy, and availability properties that their fulfillment would mitigate the identified threats. We then develop an evaluation framework based on these properties to enable an objective assessment of the protections offered by secure DNS schemes. Our work can guide the design of future schemes, by encouraging the integration of desired protections into design goals. Finally, we conduct a comparative evaluation of previously proposed DNS schemes, which aim to enhance protection without requiring fundamental changes to the original two-stage DNS resolution process. In summary, the main contributions of this paper are as follows.

\begin{itemize}

    \item We survey the threats of the DNS resolution process and develop a comprehensive threat model and attack taxonomy of this process to systematically identify and categorize existing threats (Sec.~\ref{sec:threatModel}).
    
    \item Based on the developed threat model, we define 14 security, availability, and privacy properties that their fulfillment mitigates the identified network-based threats in the name resolution process (Sec.~\ref{sec:properties}).
    
    \item We survey 12 secure DNS schemes, presenting their operational models and highlighting their security, privacy, and availability advantages and shortcomings (Sec.~\ref{sec:alternatives}).

    \item We rate the surveyed schemes using the defined properties, and we assess them to identify overarching trends in their strengths and shortcomings within the DNS resolution process (Sec.~\ref{sec:comparison}).
    
\end{itemize}
\section{Overview}
\label{sec:background}
This section provides background on the DNS resolution process, reviews related work on DNS security, and clarifies the scope and position of this survey.

\subsection{DNS Resolution Stages}
\label{sec:stages}
For efficiency, performance, and scalability reasons, the DNS resolution process typically occurs in two stages:

\subsubsection{Stage~1} As Figure~\ref{fig:resolutionpath} shows, Stage~1 is the communication path between the client (stub resolver) and the recursive resolver. Based on the configured recursive resolver on the client side, the intermediate entities in Stage~1 may vary. The client can use a recursive resolver within its local network or that of the ISP. The recursive resolver's address is typically configured automatically by the DHCP protocol if the client does not specify one manually. There are also public resolvers over the Internet, which are typically run and maintained by trusted (\ie by their users) commercial companies (\eg Google or Cloudflare). Callejo \etal~\cite{callejo2019global} in 2019 found that 13\% of the DNS clients in their study used the top 4 third-party recursive resolvers. Moura~\etal~\cite{cloudingDNS} showed that 30\% of the DNS queries to `.nz' and `.nl' country code Top-Level Domains (ccTLDs) are generated by the recursive resolvers of five top cloud providers. Lu \etal~\cite{lu2019end} also found an increasing pattern in the use of DoT and DoH with public resolvers. As such, public recursive resolvers, which are neither in the clients' local networks nor their ISPs, form a significant portion of DNS queries over the Internet. 

ASes are sets of interconnected prefixes (networks) that have a single unambiguous routing policy under control of one or multiple operators~\cite{rfc1930}. 
An ISP is typically the first AS that transfers all of the client's Internet traffic. Therefore, unlike other ASes in Stage~1 that are likely to differ based on the remote addresses that the client communicates with, the ISP is a fixed entity that transfers all of the client's Internet traffic.

\subsubsection{Stage~2} 
\begin{figure}[t!]
    \includegraphics[width=\linewidth]{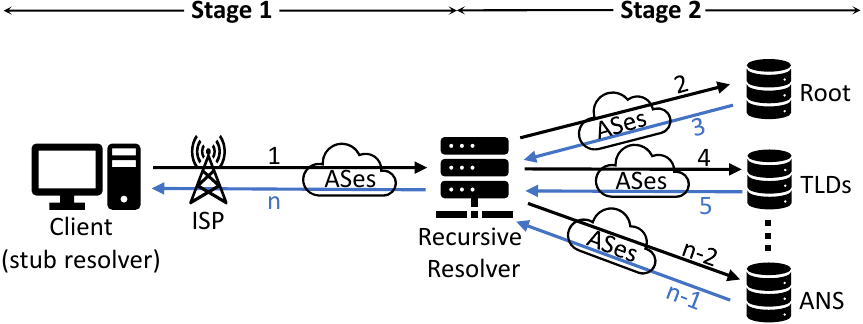}
    \caption{Complete DNS resolution path, broken into two stages: Pre-recursive-resolver (Stage~1) and Post-recursive-resolver (Stage~2).}
    \label{fig:resolutionpath}
\end{figure}

When the client’s query arrives at the recursive resolver and the Recursion Desired (RD) bit in the query is set, the resolver checks whether it has a cached answer. If no cached answer is available, Stage~2 communications between the recursive resolver and Authoritative Name Servers (ANSes) begin. In Stage~2, ANSes do not usually perform recursion themselves. If an ANS does not have the response to the client's DNS query, it will respond with the address of the next DNS server, at a lower level in the DNS hierarchy that is expected to be authoritative for the queried domain name. This procedure continues until the recursive resolver either finds the ANS for the queried domain name, or fails to find a response. As Figure~\ref{fig:resolutionpath} depicts, Stage~2 of the DNS resolution path starts from Step $2$ to Step $n-1$. Steps $n-2$ and $n-1$ are the query and the response from the ANS of the queried resource record. Stage~2 queries and responses are typically non-confidential and are transferred through multiple ASes over the Internet.

\label{ssec:ecs}
\subsection{EDNS Client Subnet (ECS)}
Typically, ANSes only see the IP addresses of recursive resolvers instead of the clients’ IP addresses. Content Delivery Networks (CDNs) aim to minimize latency by connecting clients to the closest CDN edge servers. The functionality of some CDNs requires determining a client's network address in order to provide the IP address of the closest edge server. The EDNS Client Subnet (ECS) is an extension to DNS (‘EDNS(0)’) that allows inserting a client’s IP address or subnet in queries~\cite{ecs16}. The specification of ECS suggests truncating IPv4 and IPv6 to 24 and 56 bits, to protect clients' privacy. However, this is not a strict requirement, and clients can send their 32/128 bit IPv4/v6 addresses alongside their queries~\cite{ecs16}. Even with truncation, /24 IPv4 or /56 IPv6 addresses are still likely to reveal client-related information, such as coarse-grained geographic information (\eg country, city, or organization), or is a privacy leak. Thus, a scheme that truncates ECS can still disclose client-related information that can be used by attackers (\eg for selective cache poisoning~\cite{kintis2016understanding} or censorship at the country or organization level). If a client includes ECS, an ECS-supporting recursive resolver forwards the ECS in its queries to all traversed ANSes. Consequently, all ANSes from the root and TLDs down to the authoritative name server know who is querying that domain name.

\subsection{DNS Privacy Extensions}
\textit{EDNS(0) Padding:} The EDNS(0) padding option is provided on top of the DNS extension mechanisms and allows clients and servers to add an arbitrary number of octets to mitigate size-based traffic analysis attacks~\cite{EDNS-Padding}. Although EDNS(0) padding provides privacy for encrypted DNS messages, Siby~\etal~\cite{siby2019encrypted} found that small-sized padding techniques in the secure DNS schemes do not prevent traffic analysis attacks that can reveal the clients’ visited websites and have limited impact on web browsing privacy.

\label{sec:Qmin}
\textit{QName Minimization:} Although not required, recursive resolvers typically send the full QName in their queries to ANSes when traversing the DNS hierarchy~\cite{rfc7626}. As a result, ANSes can gather information about the queried domain names from the recursive resolvers, and some queries may also include ECS information. Consequently, disclosing the full QNames reduces the privacy of clients and resolvers, allowing queried ANSes to infer information about the clients using these resolvers. For example, root receives ``www.example.com" instead of just ``.com". QName minimization was proposed to reduce the information included in queries from recursive resolvers to ANSes. Therefore, as the resolver traverses the DNS hierarchy to resolve a domain name, it includes only the necessary part of the QName in its queries.

\subsection{Previous Surveys on DNS Security and Privacy}
\label{ssec:related}
Several surveys have explored the changing landscape of DNS security and privacy. Ollmann’s early technical report on pharming~\cite{pharmingGuide} provides a detailed analysis of the DNS resolution process and architecture, identifying attack vectors within different components of the DNS infrastructure. Zou~\etal~\cite{zou2016survey} present a short summary on DNS vulnerabilities with a comparative overview of countermeasures to address them. In a comparative analysis of DNSSEC and DNSCurve, Anagnostopoulos~\etal~\cite{dnssecvs} evaluate security, privacy, deployability, and performance-related characteristics. They concluded that DNSSEC is more compatible with the core DNS protocol, while DNSCurve enhances security and privacy against active and passive on-path adversaries. Van der Toorn~\etal~\cite{VANDERTOORN2022100469} provide a tutorial and detailed examination of the DNS protocol. Their survey discusses its evolution, deployment, and modern challenges, and is aimed primarily at early-career researchers.

The survey of Kim and Reeves~\cite{kim2020survey} groups DNS security attacks into four areas: data tampering, data flooding, abuse of DNS, and vulnerabilities in DNS server structure. Privacy-related and censorship threats are beyond their scope. They briefly summarize a diverse range of mitigation techniques, including monitoring tools, malicious domain detection techniques, protocol extensions such as DANE~\cite{rfc6394} and DNSSEC~\cite{rfc4033}, and rate limiting. Their comparative evaluation is insightful, although it may be limited in its ability to capture fine-grained characteristics by the heterogeneity of the schemes. Grothoff~\etal~\cite{grothoff2018toward} survey DNS enhancements alongside radical alternatives such as Namecoin, GNS, and RAINS. They evaluate the effectiveness of these approaches against security and privacy threats including mass surveillance and DNS manipulation from inline and off-path adversaries. They also consider censorship resistance in naming systems designed to prevent even government-enforced blocking by avoiding reliance on centralized DNS operators.

Khormali~\etal~\cite{khormali2020domain} offer a comprehensive survey of DNS security and privacy threats along with prevention and detection countermeasures based on over 170 peer-reviewed publications from the ten-year period between 2010 and 2020. They also examine research methods and the impact of various DNS entities on system's security and availability. Their work does not include contributions from industry and standardization bodies such as the IETF. Schmid~\cite{SchmidThirty} provides a broad overview of the evolution of DNS security, linking major threats and protocol developments across three decades. This survey extends the historical CIA threat model with reliability to account for DNS misuse by botnets and other abuses. It also analyzes and compares six proposals aimed at securing the DNS resolution path, along with four radical alternatives to the DNS architecture.

A number of surveys have focused on techniques for detecting malicious DNS traffic, such as the use of DNS for botnet communication and data exfiltration. Lyu~\etal~\cite{lyu2022survey} present a comprehensive survey of DNS encryption protocols, including DoT, DoH, and DoQ. Their study examines adoption, performance, security benefits, and vulnerabilities of these secure DNS schemes. They highlight how encryption hinders traditional detection of malicious DNS traffic, which enables malware Command and Control (C\&C) and data exfiltration. The survey also reviews features in encrypted traffic that can be leveraged for malware detection and user activity profiling. 

Torabi~\etal~\cite{torabi2018detecting} conducted a comprehensive survey of systems that detect Internet abuse using passive DNS traffic analysis. They reviewed detection techniques across eight major systems, and highlight common design patterns, strengths, and limitations. Their analysis focused on the reliance on supervised learning and the difficulty of achieving real-time detection. To address these challenges, they proposed a solution based on big data analysis capable of near real-time performance. Similarly, Zhauniarovich~\etal~\cite{zhauniarovich2018survey} provided a comprehensive review of approaches to malicious domain detection using DNS data. Their work categorized existing methods based on evaluation metrics, data analysis techniques, and the sources of DNS data and enrichment information. Alieyan~\etal~\cite{alieyan2017survey} and Singh~\etal~\cite{singh2019issues} focused specifically on DNS-based botnet detection techniques, while Nadler~\etal~\cite{nadler2019detection} investigated methods for identifying DNS-based data exfiltration.

Compared to the above surveys, our work presents what we believe is the first comprehensive, network-based threat model and attack taxonomy that specifically targets the DNS resolution process. This taxonomy includes censorship and privacy-violating threats, which are often overlooked. In addition, we provide the first formal effort to define a set of beneficial security, privacy, and availability properties that can mitigate threats across the DNS resolution path. We also propose a systematic comparative evaluation framework for assessing secure DNS schemes. This framework highlights both the strengths and limitations of schemes that aim to enhance the protection of the DNS resolution process. Finally, we analyze twelve secure DNS schemes that augment protections in the DNS resolution path. We exclude approaches that require radical modifications to the DNS infrastructure in order to provide a balanced and fine-grained comparative analysis.
\section{Threat Model and Attack Taxonomy}
\label{sec:threatModel}

\begin{figure*}[h!]
\centering
\includegraphics[width=\linewidth]{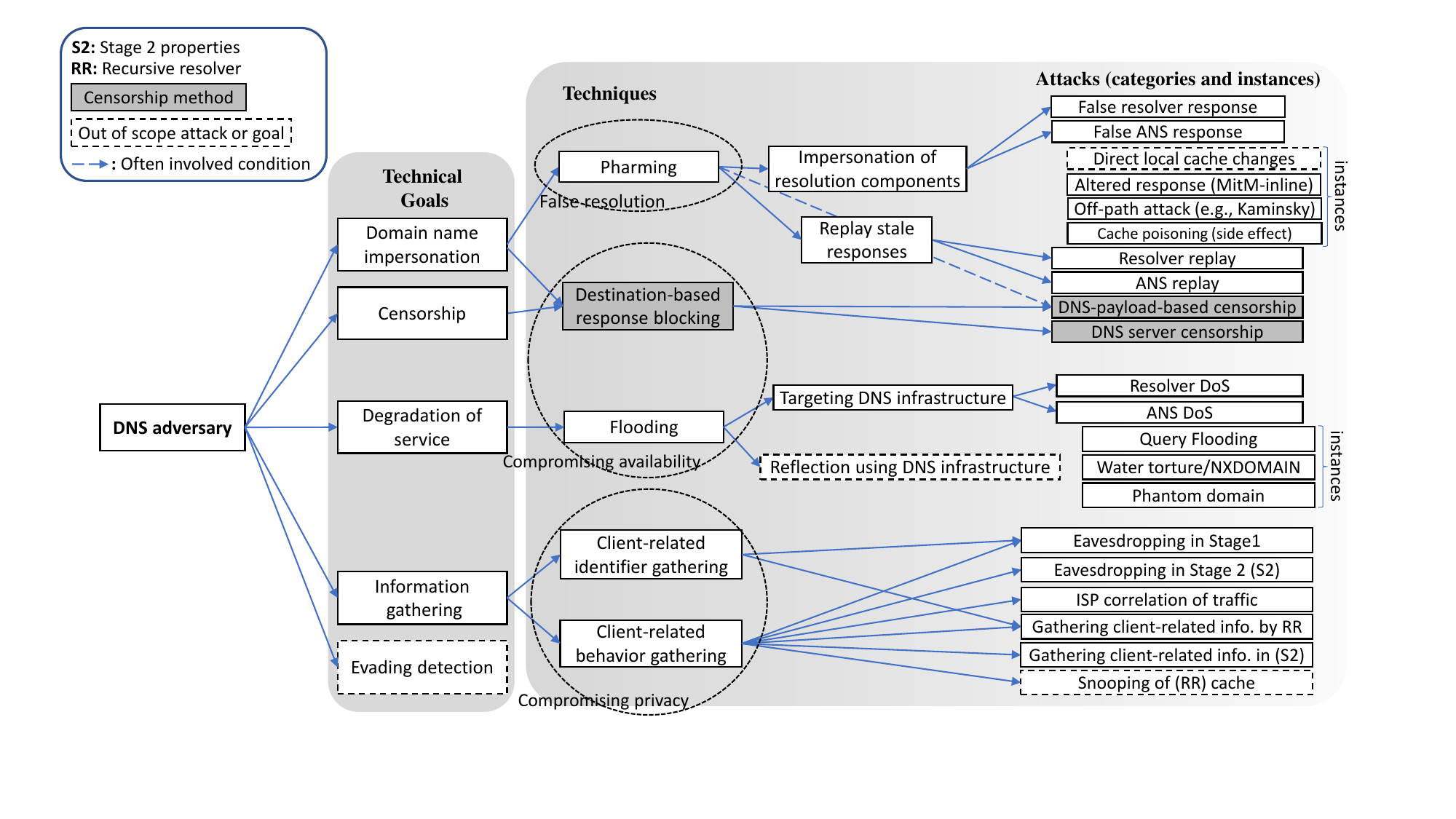}
\caption{\textbf{Threat Model and Attack Taxonomy}: Network-based attacks in DNS resolution process.}
\label{fig:ThreatTree}
\end{figure*}

We begin by defining the scope of the threats being analyzed, focusing on attacks specifically associated with the DNS resolution process. Based on the defined threat scope, we introduce a taxonomy of DNS-related threats and attack techniques. Figure~\ref{fig:ThreatTree} presents the taxonomy, classifying threats according to their technical goals and the techniques used to achieve them. Solid rectangles denote goals and techniques within our scope; dashed are out of scope.

\subsection{Threat Scope and Classification}
We define the threat model for DNS resolution, which provides the basis for analyzing the DNS schemes surveyed in this work. The DNS resolution process starts when a client generates a DNS query, and ends when the recursive resolver returns the obtained response, or error, to the client. 

Our focus is on network-based attacks in the name resolution path, which involve exploiting network protocols and specifically interfering with or affecting the name resolution process. In contrast, host-based attacks fall outside our scope. Host-based attacks typically involve a compromised host (\eg a recursive resolver infected with malware) or manipulations by a malicious insider. For example, an adversary can modify OS configurations such as the \texttt{hosts} file, primary DNS server (\eg modify network configuration files), or IP/domain-name associations on the DNS caches or databases to falsify DNS resolution. Another host-based attack targets client network configurations by using a malicious \texttt{DHCP} server to assign bogus DNS resolvers that return false responses. We do not expand on these in this work.

Additionally, attacks such as domain name squatting (\eg Typosquatting~\cite{agten2015seven}), domain name hijacking~\cite{pharmingGuide}, and parked domains~\cite{vissers2015parking} are excluded, since they are not network-based threats. Furthermore, attacks that involve registrars or search engines~\cite{pharmingGuide} are excluded because these entities are not directly involved in the name resolution process. Botnet-related DNS threats are excluded as well; these include attacks that do not interfere with the name resolution process (\eg DNS tunneling or data exfiltration~\cite{nadler2019detection}, or DNS-based communication within botnets~\cite{dietrich2011botnets}) or the attacks that exploit legitimate DNS functionality (\eg DNS-based Fast-Flux~\cite{fastfluxMatrawy} or Domain Generation Algorithms (DGAs)~\cite{DGADagon,  plohmann2016comprehensive}). Finally, DNS reflection and amplification attacks~\cite{ampattack2013dns} are excluded, as their primary target is not the availability of DNS servers themselves.

In Figure~\ref{fig:ThreatTree}, DNS threats are classified based on their technical goals and the techniques employed to achieve those goals. The main technical goals fall into five categories: domain name impersonation, censorship, degradation of service, information gathering, and evading detection. Monetization, while often viewed as the direct goal of domain parking, can be considered a broader objective that underlies many DNS attacks; however, it falls outside the scope of this paper. Monetization, while typically viewed as the direct goal of domain parking, can be considered a general objective for most of DNS attacks and falls outside the scope of this paper. 

\subsection{Domain Name Impersonation Techniques}
\emph{Domain name impersonation} refers to fraudulently representing a service point. While web services are the most common target, impersonation can also apply to non-web services such as FTP or NTP servers~\cite{jeitner2020impact}. A common technique to achieve this is pharming. One of the primary methods for domain name impersonation is pharming. From a client’s perspective, we define pharming as any method in which the DNS response (\eg the IP address in an A record) differs from the one that the authoritative name server (ANS) has currently associated with the queried record.

A common technique for performing pharming is to alter the domain name/IP address association in the cache of recursive resolvers or clients, which is an attack known as DNS cache poisoning. Cache poisoning can occur through direct local cache changes (\eg installing malware on a recursive resolver or client to manipulate cached DNS records). However, such direct local cache changes are host-based attacks and are therefore excluded from our scope.

In the network context, inline adversaries on the resolution path can inject false responses and impersonate legitimate entities to poison caches~\cite{liu2018interception}. Off-path adversaries on the Internet can achieve the same by exploiting techniques such as the Kaminsky attack~\cite{KaminskyAttack}, fragmentation-based poisoning~\cite{herzberg2013fragmentation}, network side-channels~\cite{newCP} or birthday attacks~\cite{CPandHijacking}.\footnote{The vulnerability associated with birthday attacks has been mitigated in modern DNS software implementations~\cite{afek2025pops}.} These attacks typically involve injecting false responses while impersonating legitimate recursive resolvers or ANSes, deceiving the victim into believing the responses originated from trusted sources.

Additionally, due to the absence of anti-replay protections in DNS and DNSSEC, adversaries can capture and replay stale but valid responses~\cite{ReplayStale}. For example, DNSSEC responses typically have a single signature, which is valid for a specific period (\ie between \emph{inception} and \emph{expiration} timestamps~\cite{rfc4034}). Adversaries can replay DNSSEC records that have valid signatures. For example, in instances where an ANS has reassigned a domain name to a different IP address, replaying previously signed DNSSEC records, which have non-expired signatures, can be used to misdirect clients and carry out Pharming.

\subsection{Censorship Techniques}
\emph{Censorship} involves blocking specific DNS schemes or altering DNS responses for targeted domains. We assume that governments (\eg China~\cite{GFWCensor} or Iran~\cite{aryan2013internet}), are the primary DNS-based censorship agents. In Figure~\ref{fig:ThreatTree}, techniques used for DNS-based censorship are depicted as gray boxes. \emph{Destination-based response blocking} refers to blocking DNS queries or responses that satisfy specified criteria by a censoring agent. For example, a query could include a specified domain name, a destination IP address associated with a recursive DNS server, or a destination port number associated with a particular DNS scheme. Note that direct IP-based censorship of Internet services other than DNS (\ie based on packet headers) falls outside the scope of this paper, as it is distinct from the DNS system and is not related to DNS resolution.

In \emph{DNS-payload-based censorship}, typically a Stage~1 intermediate adversary (\ie inline or on-path, such as an ISP or a recursive resolver) blocks DNS queries or responses by targeting specific payload fields. This may involve blocking or falsifying DNS resolution based on the payload's QName within a query or the IP address within a response. Since the legitimate association between a domain name and IP address often change, \emph{DNS-payload-based censorship} is a type of \emph{Pharming}.

While \emph{DNS-payload-based censorship} provides adversaries with a fine-grained controls over censorship, \emph{DNS server censorship} can block all traffic directed at the DNS servers of a specific DNS scheme based on their IP addresses, domain names~\cite{hoang2022measuring}, or their designated port number. For example, a censoring government could block DoT by blocking all traffic associated with TCP port 853.

\subsection{Degradation of Service Techniques}
\textit{Degradation of service} is another goal in which adversaries reduce or disrupt the availability of DNS resolution infrastructure for legitimate clients. We limit our primary scope to attacks where the targeted entities are the DNS resolution components. \emph{Flooding} is a technique of degrading the availability of DNS, which is broken down into attacks that \emph{target DNS infrastructure}, and attacks that perform \emph{reflection using DNS infrastructure}. 

When targeting DNS infrastructure, adversaries may flood recursive resolvers (resolver-DoS) or authoritative name servers (ANS-DoS) using techniques such as query flooding, water torture/NXDOMAIN attacks~\cite{luo2018large}, or phantom domain attacks~\cite{SchmidThirty}. In a query flooding attack, adversaries send a large volume of DNS queries to a UDP-based recursive resolver or ANS in order to exhaust its resources, primarily the CPU DNS servers. A variation of this is the water torture/NXDOMAIN attack, where adversaries generate random strings and append them as prefixes (\ie subdomains) to queries~\cite{luo2018large}. Since the targeted ANS must attempt to resolve these non-existent domains, the attack depletes ANS resources while also filling the recursive resolver’s cache with non-existent responses.

In a \emph{phantom domain} attack, the adversary establishes slow or unresponsive ANSes (\ie phantom domains) and then directs queries for these domains toward a victim recursive resolver. Because the resolver must repeatedly interact with these unresponsive ANSes, its resources are consumed, degrading its performance.

In reflection attacks using DNS infrastructure, adversaries exploit the stateless nature of UDP-based DNS to reflect and amplify flooding traffic, thereby obscuring the source of the attack. These reflection attacks are excluded from our scope, as their primary goal is not to disrupt DNS availability directly but to amplify general DoS traffic.

\subsection{Information Gathering Techniques}
\emph{Information gathering} refers to adversaries collecting client and query-related information along the name resolution path, often violating client privacy.
As entities involved in the name resolution process typically have access to the data and metadata of transmitted DNS messages (\eg query payload or metadata), they can gather various types of information about clients, thereby violating client privacy. \textit{Information Gathering} is the third goal where adversaries (typically intermediate entities in the name resolution process) collect information on client behavior and identity using DNS queries, responses, and their associated metadata. For instance, collecting a client’s browsing or DNS query history and combining it with additional information has been shown to enable identification or re-identification of the client across different time periods~\cite{Birdreidentifiability, narayanan2010myths}.

Domain names within DNS queries may disclose various information with different privacy sensitivity levels. For example, DNS queries can leak personal details about a client, such as the applications they use or the types of IoT devices they own~\cite{apthorpe2017closing}. In some cases, DNS queries may be directly linkable to a real-world entity, making the leaked data personally identifiable information (PII). For example, if a client issues a DNS query with the QName `\emph{admin.example.com},' it is probable that the client is the administrator of `\emph{example.com},' thus linking the query to a real-world identity.

In the DNS context, some part of the payload can directly disclose client-related behavior and some part of the metadata (transmission-related information) can be used for client identification. For example, the Questions, Answers, Authority, and Additional sections of a DNS message payload can directly reveal client-related behavioral information. However, DNS headers and other network layer headers can be collected to identify a client or infer behavior over time. 

In Figure~\ref{fig:ThreatTree}, \textit{client-related behavior gathering} refers to collecting information from DNS queries that directly reveal client activities, while \textit{client-related identifier gathering} refers to collecting metadata or payload elements that could serve as direct (\eg IP addresses) or indirect (\eg timestamps) identifiers. Indirect identifiers can also be combined with other information to improve identification accuracy or lead to the creation of PII~\cite{narayanan2010myths}.

\textit{Eavesdropping} can occur in both stages. In Stage~1, an \textit{Eavesdropping} can occur in both stages of DNS resolution. In Stage~1, an adversary can collect both client-related behavioral patterns and identifying information (\ie IP address) from a DNS query by \emph{eavesdropping}. In Stage~2, adversaries can collect client-related behavioral patterns through \textit{Eavesdropping}, while the metadata (\ie identifying information) belong to the recursive resolver. Although metadata in Stage~2 does not directly belong to clients, the DNS query payload may contain client-related identifiers such as ECS (see Sec.~\ref{ssec:ecs}). In Figure~\ref{fig:ThreatTree}, we assume that there is no ECS included in Stage~2; thus, \emph{eavesdropping in Stage~2} only reveals client-related behavior. 

In Stage~1, the ISP acts as the convergence point for the client’s Internet interactions. As such, it can correlate encrypted DNS queries with subsequent client traffic, enabling discovery of the encrypted queried DNS record. For example, if a client sends an encrypted DNS query followed by HTTPS traffic, the ISP can perform a reverse lookup or inspect the Server Name Indication (SNI) field in the TLS \texttt{ClientHello} message to infer the QName of the earlier encrypted DNS query. In Figure~\ref{fig:ThreatTree}, this is shown as ISP correlation of traffic, an attack against the confidentiality of encrypted DNS that reveals client-related behavior.

Beyond ISPs, recursive resolvers can also collect information from the DNS queries they receive.\footnote{\eg For example, Google's public DNS servers' logged information: https://developers.google.com/speed/public-dns/privacy} In \emph{gathering client-related information by (RR)}, a recursive resolver can log both a client's behavior-related (\eg QName) and identity-related (\eg IP address) information. Recursive resolvers can also geolocate clients based on their IP addresses. In \mbox{\textit{gathering client-related information (S2)}}, the query metadata primarily belongs to the recursive resolver, while the payload may still reveal client-related information (including ECS, if present). Therefore, different entities involved in Stage~2 (\ie intermediate ASes and ANSes) can infer information about the behavior of clients that use a recursive resolver but cannot reliably link it to individual clients unless ECS is included.

Finally, DNS \textit{cache snooping} is an active information-gathering technique in which attackers infer previously resolved domain names by probing a recursive resolver’s cache. \textit{Cache Snooping} is often achieved by sending non-recursive queries (RD=0), inspecting the remaining Time To Live (TTL) in a recursive resolver's responses, or measuring the response time of the DNS cache~\cite{grangeia2004dns}. However, \textit{Cache Snooping} against large, multi-layered centralized resolvers with distributed caches is more sophisticated than probing single-cache resolvers~\cite{trufflehunter}. Since cache snooping relies on legitimate DNS functionality (\eg non-recursive requests) or timing side channels (\eg latency analysis); it is considered out of scope for this paper.

\subsection{Evading Detection Techniques}
As nearly all Internet-connected networks rely on DNS for name resolution, DNS traffic is rarely blocked by firewalls. In \emph{evading detection} based on DNS, adversaries exploit DNS as a benign protocol to conceal malicious traffic or infrastructure. For example, botnets may use DGAs and DNS-based fast-flux~\cite{fastfluxMatrawy} to hide their C\&C server~\cite{alieyan2017survey, plohmann2016comprehensive}. Moreover, attackers can exfiltrate arbitrary data from compromised targets via DNS tunneling. Unlike other forms of attack, DNS-based evasion does not directly interfere with or disrupt the name resolution process; rather, it leverages the legitimate functionality of DNS for malicious purposes. For this reason, the technical goal of evading detection is considered out of scope in this paper.
\section{DNS Scheme Properties}
\label{sec:properties}
Before evaluating DNS schemes, we define 14 properties grouped into three categories: security, availability, and privacy. Numerous secure DNS schemes have been proposed in the literature to improve one or more of these aspects, and many share operational similarities. To provide a consistent basis for comparison, this section defines the DNS resolution-related properties used in our analysis. In the next section, we survey secure DNS schemes, discussing their theory of operation and security design goals in terms of the defined properties.

\subsection{Security Properties}
\begin{itemize}[leftmargin=1.5em]
\label{ssec:secProps}
    \item[\textbf{S1}] \hypertarget{S1}{\textbf{\SResImp}}: \defin{This property is achieved if a DNS scheme prevents adversaries from injecting false responses to clients by impersonating legitimate recursive resolvers.} In Stage 1, if an adversary can observe DNS queries issued by a client or can trigger the client to generate queries (\eg via embedded images on a webpage), it can attempt to inject false responses by impersonating the legitimate queried recursive resolver. Impersonation involves generating a false response that uses parameters from the client’s query (\eg the resolver’s IP address) so that the response appears authentic from the legitimate recursive rsolver. An adversary in Stage 1 may be an inline or intermediate entity (\eg a rogue router or ISP) with direct access to queries, or an off-path adversary that redirects or observes traffic through hijacking techniques (\eg ARP spoofing or BGP hijacking). 

    Inline adversaries can directly block or modify queries and responses, while off-path adversaries need to trigger the target client to issue DNS queries and must guess query parameters (\eg transaction ID, source port) and inject their false response before the legitimate response arrives. Without message authentication, such injections would go undetected, enabling adversaries to impersonate resolvers and potentially poison the client’s cache (\eg stub resolvers or browser caches). Moreover, by providing message authentication, an scheme implicitly ensures data integrity~\cite{hac1996}.
	
    \item[\textbf{S2}] \hypertarget{S2}{\textbf{\SResReplay}}: \defin{A DNS scheme provides this property if responses from a recursive resolver cannot be replayed to other clients, nor can old responses be replayed later to the same client.} One example of anti-replay means is employing authenticated encryption with fresh session keys per client-resolver interaction prevents the replay of responses across clients or across sessions of the same client. Moreover, incorporating Time-Variant Parameters (TVPs) [64] into protected DNS messages mitigates replay within a session. In the absence of such anti-replay mechanisms, adversaries can replay stale responses to misdirect clients (\eg Pharming) or to degrade services (\eg CDN availability~\cite{hao2018end}).
    
    \item[\textbf{S3}] \hypertarget{S3}{\textbf{\SANSImp}}: \defin{A DNS scheme satisfies this property if adversaries cannot successfully inject false responses to recursive resolvers by impersonating legitimate ANSes.} Without this property, an adversary can impersonate ANSes used in recursive resolver queries and inject forged responses, leading to cache poisoning and Pharming. Since open resolvers accept queries from any IP address, both inline and off-path adversaries can trigger them to issue queries and attempt false response injection. Providing message authentication in Stage~2, where recursive resolvers validate responses from ANSes, is one way to achieve this property. Unlike Stage~1 false response injections, which typically affect only a single client, successful ANS response injection in Stage~2 compromises a recursive resolver’s cache and can thereby impact many clients that rely on it.
    
    \item[\textbf{S4}] \hypertarget{S4}{\textbf{\SANSReplay}}: \defin{A DNS scheme satisfies this property if responses from ANSes cannot be replayed either to other recursive resolvers or later to the same recursive resolver.} Without this property, adversaries can capture and replay stale responses, which may no longer be correct. For example, in DNSSEC, although responses are signed, they can be replayed by entities other than ANSes until their signatures expire~\cite{yan2008limiting}. Such replayed records can inject false entries into resolver caches~\cite{ReplayStale}, misdirecting clients to obsoleted IP addresses. One way to mitigate this threat is for ANSes to use distinct session keys for authenticating responses to different recursive resolvers in Stage~2, ensuring uniqueness of each interaction. Additionally, incorporating TVPs into exchanged protected DNS messages prevents the reuse of previously authenticated responses, even when the same key is used for securing multiple messages with a resolver.

    \item[\textbf{S5}] \hypertarget{S5}{\textbf{\SKeyExp}}: \defin{This property is satisfied by a DNS scheme that prevents the duplication of long-term private keys across server instances within a zone, thereby reducing the risk of key exposure in the event of server compromise.} Long-term private keys are often used in cryptographic operations such as signing to provide security guarantees in Stage~2. However, replicating these keys across multiple ANS instances increases the likelihood of compromise. Moreover, some servers may be located in countries or jurisdictions where a zone owner may not prefer to duplicate long-term keys in those locations. For example, in IP anycast deployments, storing the same long-term private key on all CDN servers exposes it to compromise if any single server is compromised. By contrast, in DNSSEC, signatures are pre-computed, allowing the long-term private key to be stored securely in one location without needing to be present on all ANS servers.

\end{itemize}

\subsection{Availability Properties}
\begin{itemize}[leftmargin=1.5em]
\label{ssec:availProps}
    \item[\textbf{A1}] \hypertarget{A1}{\textbf{\SECensor}}: \defin{A DNS scheme must prevent a Stage~1 adversary from gaining access to plaintext DNS query and response sections to satisfy this property. Consequently, an attacker cannot alter name resolution based on specified DNS payload sections to block the resolution process.} In Stage~1, censoring countries, such as China~\cite{GFWCensor} or Iran~\cite{aryan2013internet}, often leverage DNS to block access to specific domain names. Various sections in DNS queries and responses can serve as the basis for censorship patterns. The QName field inside the Question section is the one most often targeted by censoring agents. In this paper, if a DNS scheme prevents censoring agents from accessing the Question, Answer, Authority, and Additional sections of the DNS payload, it satisfies this property.\footnote{Some public recursive resolvers remove Authority or Additional sections from DNS responses. We assume that these sections are not stripped from DNS responses since this behavior is not consistent across all resolvers.} As the DNS message header section contains only non-privacy-sensitive information (\eg control flags), a DNS scheme that allows this section to remain visible while concealing the other sections is still considered to satisfy the property. Inline censoring entities in Stage~1 (\eg ISP) can target specific DNS queries or responses based on one or more DNS payload sections. These targeted DNS queries or responses can then be blocked or replaced with an NXDOMAIN answer~\cite{censorIndia}, a routable public IP address (\eg Facebook address~\cite{GFWCensor}), or a false IP address (\eg a placeholder “black hole” address~\cite{aryan2013internet}).

    \SECensor{} is typically achieved by encrypting queries and responses in Stage~1. Strongly obfuscating DNS messages is another technique to achieve this property. Our rationale for defining this property only in Stage~1 is that an informed user or privacy-aware Internet software (\eg web browsers) can choose recursive resolvers located outside censoring regions. Thus, if a DNS scheme provides \SECensor{} in Stage~1, the recursive resolver could be selected in a non-censoring region and the property is not required in Stage~2 to mitigate censorship. For example, a client who uses Cloudflare’s recursive resolver assumes that the resolver is not controlled by a censoring entity and is located in a non-censoring region. Therefore, if queries from the client to the resolver are \SECensor{}, the property is not required in Stage~2.

    \item[\textbf{A2}] \hypertarget{A2}{\textbf{\ServerCensor}}: \defin{A DNS scheme provides this property if clients' interactions with recursive resolvers do not have distinguishable characteristics that can be used to block access to those recursive resolvers.} Compared to \hyperlink{A1}{A1}, this is a more coarse-grained censorship technique, where a censoring entity completely blocks access to the recursive resolvers of a DNS scheme. Inline middleboxes under the control of a censoring government can be used to block DNS requests or responses with specific characteristics. For instance, access to the servers of a DNS scheme can be blocked by their IP addresses, domain names, or port numbers. If a DNS scheme is limited to a set of discoverable IP addresses or domain names, or if it uses a distinguishable port number, these characteristics can be used to censor the scheme by blocking access to its resolvers. Additionally, if a DNS scheme uses a known port number that directly indicates its use (\eg TCP/853 for DoT), the IP addresses of its servers can be effectively discovered over the Internet (\eg using ZMap~\cite{zmapscanner}). Thus, a DNS scheme that relies on a distinguishable port number can be censored either directly by port blocking or indirectly by identifying its server IP addresses and blocking access to them.\footnote{If the IP addresses of the DNS servers are frequently changed, persistent scanning and censoring can still effectively block access to these servers.} Moreover, if a DNS scheme is built on top of another protocol that itself can be censored, then blocking the underlying protocol effectively censors the DNS scheme as well. For example, a censoring government can block access to Tor relays or directory authorities to disrupt DNS-over-Tor resolvers.
    
    One way to achieve \ServerCensor{} is to design a DNS scheme that does not depend on another censorable protocol, does not have distinguishable characteristics (\eg unique port number), and is not limited to a set of easily discoverable IP addresses or domain names. DNS schemes that disguise their traffic using ports associated with protocols tend to be more resilient to port-based censorship. In such cases, censoring traffic requires fingerprinting and traffic analysis, which is more costly. For instance, DoH traffic uses port 443 (shared with HTTPS), so censoring DoH would be costly (albeit not impossible). Schemes whose traffic can still be effectively identified and isolated through additional analysis techniques (\eg machine learning or deep learning classification techniques or statistical analysis) receive half credit for \hyperlink{A2}{A2}.
	
    \item[\textbf{A3}] \hypertarget{A3}{\textbf{\AResDoS}}:\defin{A DNS scheme provides this property if the transport layer protocol used in Stage~1 offers resistance mechanisms against DoS attacks, or if the scheme employs application-layer DoS mitigation techniques to protect recursive resolvers.} To ensure the availability of the DNS resolution process, recursive resolvers must remain accessible to legitimate clients and resist DoS attacks. Although perfect availability on the Internet is idealistic, some DNS schemes are more resilient to DoS attacks than others. The primary type of attack against availability is Distributed Denial of Service (DDoS), which overwhelms the resources of a target from multiple locations. Bandwidth depletion is a general type of DoS that can target any Internet-connected service, including recursive resolvers. However, DNS-related DoS attacks may specifically target DNS at the application layer (\eg flooding in UDP-based DNS) or the underlying transport protocols upon (\eg TCP SYN flooding in TCP-based schemes).
    
    As a UDP-based scheme, Vanilla DNS does not verify the source address of queries at the transport layer, and queries are directly answered at the application layer. During a DDoS attack, as the query volume increases, server CPU usage, response latency, and the number of unanswered queries also increase~\cite{zhu2015TDNS}. Furthermore, if source address spoofing is employed in UDP-based DNS packets, query rate-limiting based on IP addresses becomes ineffective~\cite{zhu2015TDNS}. In contrast, DDoS attacks on TCP-based DNS schemes often rely on overwhelming pre-allocated resources for TCP connections through SYN flooding, preventing legitimate users from accessing name resolution services~\cite{van2020computer}. However, unlike UDP, TCP-based DNS does not serve queries at the application layer before the handshake is completed, making IP-based rate-limiting a viable defense after the handshake. Additionally, TCP-based DNS can employ SYN cookies and SYN cache~\cite{lemon2002resisting} to mitigate SYN flooding attacks~\cite{zhu2015TDNS}.
    
    In summary, if a DNS scheme operates over TCP, it is generally considered resilient against DoS attacks, as TCP provides mechanisms such as SYN cache and SYN cookies~\cite{lemon2002resisting}), and queries are not served at the application layer until the handshake is completed. On the other hand, due to the lack of mechanisms for detecting spoofed messages, UDP-based DNS schemes respond to spoofed queries at the application layer and are therefore more susceptible to DoS attacks. To strengthen UDP-based schemes, anti-DoS mechanisms can be implemented at the application layer (\eg as in the QUIC protocol).

    \item[\textbf{A4}] \hypertarget{A4}{\textbf{\AANSDoS}}: \defin{Similar to \hyperlink{A3}{A3}, a DNS scheme satisfies this property if ANSes are provided with DoS resistance mechanisms at the transport layer or if the scheme employs DoS mitigation techniques at the application layer.} In addition to recursive resolvers, ANSes are necessary components for a successful DNS resolution process, and they are also susceptible to DoS attacks. Therefore, we apply the same rule as in \hyperlink{A3}{A3} when evaluating resilience of ANSes to DoS. Specifically, in Stage~2, TCP-based schemes and UDP-based protocols that employ application-layer DoS mitigation techniques are considered to provide \AANSDoS.
\end{itemize}

\subsection{Privacy and Anonymity Properties}
\begin{itemize}[leftmargin=1.5em]
\label{sec:privacyprops}
    \item[\textbf{P1}] \hypertarget{P1}{\textbf{\PEavesOne}}:  \defin{A scheme satisfies this property if the plaintext content of DNS messages (at the application layer) transmitted in Stage~1 is accessible only to authorized entities (\ie clients and recursive resolvers).} Since Vanilla DNS lacks confidentiality, queries and responses in Stage~1 are susceptible to eavesdropping. Despite being a security property, we categorize confidentiality of DNS queries and responses as a privacy property. Our rationale is that DNS information is not secret; it is inherently public. Nevertheless, transmitting DNS in plaintext compromises clients’ privacy.

    Encrypting the DNS messages transmitted from stub to recursive is one approach to achieve \PEavesOne. In this case, inline entities such as a client's local network, ISP, or other Stage~1 ASes cannot access the content of DNS messages. However, if the used encryption algorithm cannot resist traffic analysis attacks that reveal queried domain names, the scheme does not receive full credit for this property. In evaluating this property, it is assumed that the traffic of a DNS scheme is isolated and then assess its resistance to eavesdropping. Therefore, this property is evaluated independently from the rating of \hyperlink{P2}{P2}.
    
    \item[\textbf{P2}] \hypertarget{P2}{\textbf{\PConceal}}: \defin{To satisfy this property, detecting and isolating a DNS scheme's traffic by means of explicit headers or behavioral characteristics have to be prevented in Stage~1.} If DNS traffic is distinguishable in Stage~1, intermediate entities can block, redirect, or censor it, and in cases requiring backward compatibility, force the communication to fall back~\cite{downgradeDoH} to less secure schemes (\eg Vanilla DNS).
    
    Furthermore, if the traffic of an encrypted DNS scheme is detectable, it can get filtered and be subjected to traffic analysis (\eg in DoT/DoH~\cite{siby2019encrypted}) to infer the visited websites by clients. In Stage~1, if the traffic of a DNS scheme is not easily distinguishable by explicit headers in the network or transport layer (\eg destination port number), the scheme partially provides \PConceal. For instance, some DNS schemes, such as DoT, use a specific destination port number (853), which renders their traffic easily distinguishable and therefore prevents them from satisfying this property.
    
    In addition to explicit headers, if the traffic of a DNS scheme is also not distinguishable by behavioral characteristics such as timing or size analysis, the scheme receives full credit for this property. To resist behavioral analysis, a DNS scheme may use techniques such as padding before encryption or traffic-flow security~\cite{thakur2016connectivity} (\eg sending cover traffic) to render its traffic less distinguishable. For example, DoH merges its traffic with the web by using port 443, transferring DNS queries as HTTPS traffic. However, although DoH traffic is not distinguishable by its explicit headers, it remains susceptible to behavioral analysis (\eg packet size~\cite{dohisolate}, connection duration~\cite{vekshin2020doh}) that can isolate DoH traffic. Thus, DoH partially provides conceals DNS message nature.
	
    \item[\textbf{P3}] \hypertarget{P3}{\textbf{\PDecouple}}: \defin{A DNS scheme satisfies this property if it provides any means to conceal a client's IP address from the queried recursive resolvers in Stage~1.} Although DNS message confidentiality in Stage~1 (see \hyperlink{P1}{P1}) prevents unauthorized access by inline entities to DNS queries, recursive resolvers typically have access to both client-related behavior (\eg QName) and identifying information (\eg IP address). In Stage~1, recursive resolvers can collect query-related information and then use it to analyze the behavior of DNS clients.
    
    For example, Bird~\etal~\cite{Birdreidentifiability} showed that clients can be accurately and uniquely identified over different periods based on their web browsing histories. The same type of identification and re-identification can be achieved by collecting DNS-related data from clients. Moreover, in the context of IoT devices, the DNS query payloads and metadata can disclose information about the types of IoT devices and the applications used by a clients to entities involved in Stage~1 of the resolution process (\eg recursive resolver)~\cite{apthorpe2017closing}.
    
    Some DNS schemes use techniques such as proxies, relay servers, or the Tor network to enhance client privacy against resolvers, thereby separating a client's IP address from its DNS queries. However, if a client (stub resolver) includes its IP address or subnet as ECS in DNS queries, the anonymizing DNS scheme adds overhead without preserving privacy. Therefore, schemes that hide clients' IP addresses from recursive resolvers must explicitly remove or truncate ECS from their queries. 

    \item[\textbf{P4}] \hypertarget{P4}{\textbf{\PEavesTwo}}: \defin{Similar to \hyperlink{P1}{P1}, this property is satisfied if the plaintext content of application-layer DNS messages in Stage~2 is accessible only to authorized entities (\ie recursive resolvers and ANSes).} If the traffic between recursive resolvers and ANSes (\ie Stage~2) is confidential, we consider a DNS scheme \PEavesTwo{}. Consequently inline entities cannot access DNS message contents. Encryption is typically used to preserve the confidentiality of DNS messages in Stage~2.
    
    An inline adversary in Stage~2 may still be able to infer the QName of a recursive resolver's encrypted query through inter-domain dependencies, the set of name servers queried by the recursive resolver~\cite{hayaBadPrivacy, ramasubramanian2005perils}, or other side channels. However, confidentiality in Stage~2 still prevents inline entities from accessing to client-related information even if ECS is included in the queries. If the used encryption algorithm is vulnerable to traffic analysis attacks that can reveal queried domain names, the scheme is not fully credited for this property. We do not further pursue the discussion of end-to-end encryption between clients and ANSes because it noticeably increases the query overhead of ANSes~\cite{hayaBadPrivacy}, and bypasses the benefits, which are achieved with using caching recursive resolvers.
 
    \item[\textbf{P5}] \hypertarget{P5}{\textbf{\PHidesTwo}}: \defin{A DNS scheme that removes, truncates, or conceals ECS in DNS queries to preserve client anonymity in Stage~2 satisfies \PHidesTwo{}}. Two types of entities can compromise client privacy in Stage~2: ANSes queried by a recursive resolver and intermediate entities between a recursive resolver and ANSes (when they can observe the plaintext of queries). 
    
    If a client includes ECS in DNS queries, recursive resolvers that support ECS will also include ECS in their iterative queries as they traverse the DNS hierarchy. Therefore, the intermediate entities in Stage~2 and the ANSes (from root servers to the authoritative name server of the queried domain) will have access to the ECS added by the client. If the exact IP address of a client is included (\ie /32 in IPv4 or 128 bits in IPv6), it can be used for user tracking, information gathering, selective cache poisoning~\cite{kintis2016understanding}, or ECS-based censoring.
    
    When ECS is truncated, for example by using the default recommendation in the ECS RFC of 24 bits for IPv4 and 56 bits for IPv6, the resulting truncated addresses may still reveal client-related information (\eg country, city, or organization). Thus, a scheme that truncates ECS can still disclose client-related information that attackers may use, such as for selective cache poisoning~\cite{kintis2016understanding} at the country or organization level. Such schemes therefore receive only partial credit for the \PHidesTwo{} property. If a DNS scheme eliminates ECS in DNS queries, thereby preventing Stage~2 intermediate entities and ANSes from accessing client IP addresses or subnets, it completely provides the \PHidesTwo{} property. Completely removing ECS improves both the privacy and security of clients, and may also reduce page load times by avoiding the overhead caused by ECS record cache misses~\cite{hounsel2020comparing}.
    
    Additionally, DNS schemes that encrypt queries in Stage~2, and thereby preventing Stage~2 intermediate entities from accessing ECS, receive partial credit for this property. In such schemes, intermediate entities cannot access ECS, but ANSes still have access to the included ECS. If clients do not include ECS in DNS queries, no client-related identifiers are  disclosed in Stage~2. In our evaluation, we assume that ECS is included in DNS queries. A DNS scheme must explicitly remove, encrypt, or truncate ECS at stub or recursive resolvers in order to satisfy \PHidesTwo{}.
\end{itemize}

\section{Summary of Secure-DNS Alternatives}
\label{sec:alternatives}
In this section, we survey and analyze the security, privacy, and availability properties provided by 12 DNS schemes proposed to secure the DNS resolution process. Each scheme is classified into three categories based on the stage in which it is commonly applied in practice. Some schemes are designed to secure Stage~1, others Stage~2, while a small subset targets both stages.

\subsection{Rating Vanilla DNS}
\begin{figure}[h!]
\centering
\includegraphics[width=0.9\linewidth]{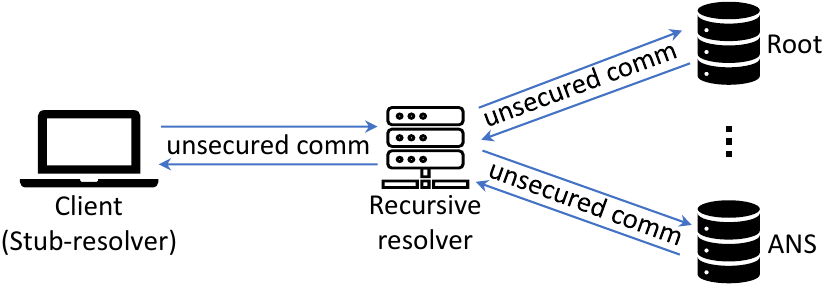}
\caption{\textbf{Vanilla DNS}: No security, privacy, or availability properties.}
\label{fig:vanilladns}
\end{figure}
Originally, DNS was designed with no security, and none of the security properties defined in Section~\ref{sec:properties} had been considered. Vanilla DNS relies on UDP transport protocol (except for the responses that exceed the defined maximum size), and does not provide any mechanisms to mitigate DoS attacks at the application layer. As a result, adversaries can overwhelm recursive resolvers and ANSes using spoofed queries, and Vanilla DNS is not resilient to DoS in both stages.

\subsection{S-DNS}
\begin{figure}[h!]
\centering
\includegraphics[width=0.9\linewidth]{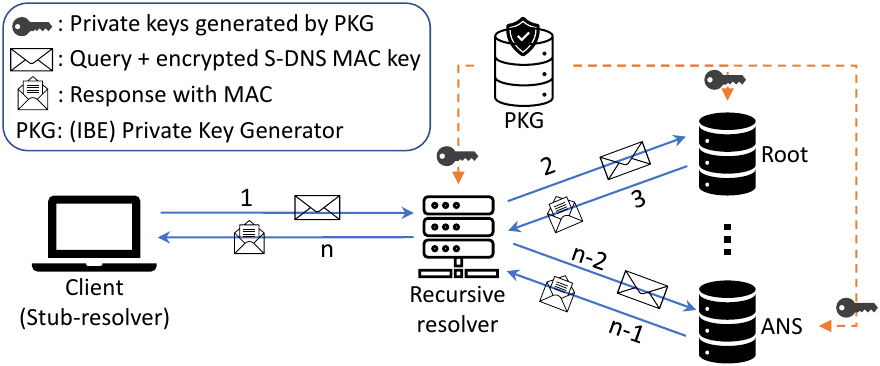}
\caption{\textbf{S-DNS}: Employing IBE-based encryption and MAC to prevent false response injections.}
\label{fig:SDNS}
\end{figure}

Bassil~\etal~\cite{bassil2012security} proposed S-DNS as a backward-compatible scheme to mitigate cache poisoning attacks. They briefly described how recursive resolvers and ANSes can use Identity-Based Encryption (IBE) to add message authentication by including a Message Authentication Code (MAC) in the additional section of DNS responses. As illustrated in Figure~\ref{fig:SDNS}, a trusted IBE Private Key Generator (PKG) is responsible for generating private keys for DNS responders (\eg recursive resolvers and ANSes) within the DNS ecosystem. A requester initiating a DNS query (\eg Steps 1, 2, or n–2 in Figure~\ref{fig:SDNS}) generates a 128-bit random bit string, referred to as the S-DNS secret, which serves as a MAC key. This secret is then encrypted using the responder’s IBE-derived public key and appended to the additional section of the DNS query. Upon receiving the query, the responder decrypts the S-DNS secret and uses it to compute a MAC for the DNS response (\eg in Steps 3, n–1, n). When the response is received, the requester recomputes the MAC using the previously generated S-DNS secret and verifies it against the received MAC, thereby ensuring the integrity and authenticity of the DNS response.

S-DNS is designed to provide security properties for the communications in both stages. By adding MACs to DNS responses, it provides \SANSImp{} and \SResImp{}. Furthermore, when the S-DNS secret is used once, the scheme also satisfies \SResReplay{} and \SANSReplay{}. However, all zones are required to trust the PKG, as it is responsible for generating and managing the private keys of each zone. If the PKG is compromised or goes rogue, the security of all zones relying on PKG-generated private keys is undermined. In addition, a central PKG does not scale well for large systems or global deployment, and a secure channel is required for transmitting private keys from the PKG to ANSes and resolvers.

\subsection{DNSSEC}
\begin{figure}[h!]
\centering
\includegraphics[width=\linewidth]{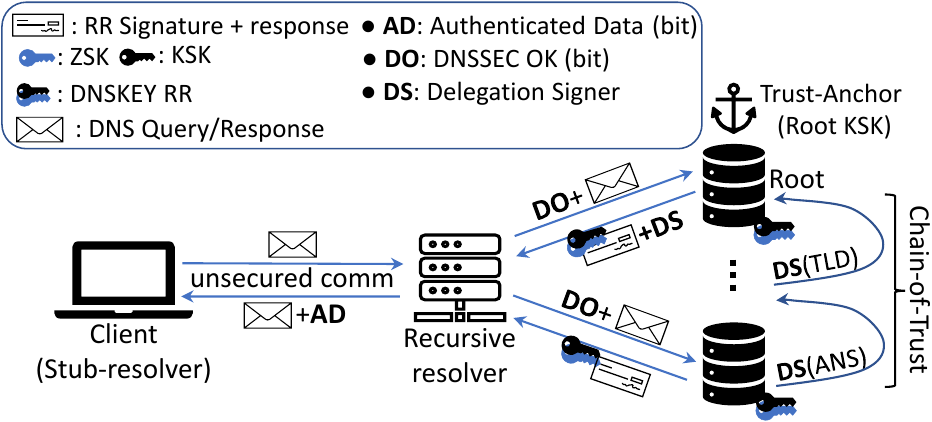}
\caption{\textbf{DNSSEC}: Using a trust anchor and signed records to provide data origin authentication for DNS records.}
\label{fig:dnssec}
\end{figure}

Based on the DNS Security Extensions (DNSSEC) specifications (\ie Requests For Comments (RFCs) 4033-4035~\cite{rfc4033,rfc4034,rfc4035}), DNSSEC (as depicted in Figure~\ref{fig:dnssec}) extends Vanilla DNS with security mechanisms that provide message authentication, integrity, and authenticated denial of existence. DNSSEC leverages public key cryptography by introducing two public/private key pairs to each DNS zone: the Key-Signing-Key (KSK) and the Zone-Signing-Key (ZSK)~\cite{rfc4033}. The private part of the ZSK is used to sign the Resource Record (RR) sets with the same name, type, and class in a zone, except for DNSKEY records. DNSKEY is a DNSSEC-specific RR that contains the public keys used in the authentication process, and this record is signed by the private part of the KSK. A hash of a DNSKEY record (\ie the public part of the KSK) known as the Delegation Signer (DS) is sent to the parent zone in a delegation. The parent zone signs and stores the DS record, creating a chain of trust from the delegated zone up to the root. Therefore, DNSSEC-enabled zones rely on their superordinate zones to properly establish the chain of trust up to the root. At the root of this hierarchy, the public KSK of the root zone acts as the trust anchor and is pre-installed in resolver software.

As Figure~\ref{fig:dnssec} illustrates, a \emph{non-validating} \emph{security-aware} client (\ie a stub-resolver) sends a recursive query (RD bit set) to a recursive resolver with the DNSSEC OK (DO) bit set. A \emph{non-validating} \emph{security-aware} stub resolver understands DNSSEC-related extensions but does not validate signatures on its own~\cite{rfc4033}. Instead, it relies on a trusted recursive resolver upstream in the DNS resolution path to verify DNSSEC keys and signatures. The DO bit indicates that the client is DNSSEC-aware, thus, the DNS server can include DNSSSEC-related data (\eg records or headers) in its response~\cite{rfc4033}. If the recursive resolver supports DNSSEC, when the resolver queries the ANSes, in addition to the queried record, it receives DNSSEC records (\eg signatures, DS, and DNSKEY) from the ANSes. The recursive resolver then validates DNSKEY records of child zones using the signed DS records received from their parent zone. It subsequently validates RR Signatures (RRSIGs) using the verified keys from the DNSKEY records. If the entire validation chain is successful, the resolver sets the Authenticated Data (AD) bit and sends the response to the stub-resolver (client); otherwise, the AD bit is left unset.

In this procedure, although the stub-resolver is ``security-aware'' (meaning it checks the AD bit), it trusts the recursive resolver and the network over which it transfers the DNS messages (\ie Stage~1)~\cite{rfc4033}. In DNSSEC, security-aware stub resolvers can also be \textit{validating}. These resolvers validate the signatures and keys voluntarily instead of relying on the recursive resolver and the inline network entities in Stage~1. However, to enhance compatibility and reduce complexity and overhead (on both clients and ANSes), stub resolvers are often implemented as non-validating but security-aware (\eg in Windows~7 and later OSes\footnote{\url{https://learn.microsoft.com/en-us/previous-versions/windows/it-pro/windows-server-2012-r2-and-2012/dn593685(v=ws.11)}}). As a result, Stage~1 communication in DNSSEC remains insecure, and intermediate MitM adversaries can send false responses marked as DNSSEC-verified responses (\ie AD bit set) to clients.

DNSSEC does not provide properties for Stage~1, leaving it vulnerable to attacks between clients and recursive resolvers. However, DNSSEC provides \SANSImp{} in Stage~2 by validating RRSIGs using DNSSEC’s chain of trust. Furthermore, since RRSIGs are pre-computed for records within a zone, the long-term private key can be stored securely without requiring DNSSEC keys to be available on all name server instances of a zone. Hence, DNSSEC satisfies the \SKeyExp{} property. Due to the lack of real-time signing, anyone can replay signed DNSSEC messages as long as their signatures are valid. Thus, DNSSEC is not \SANSReplay{}, and an adversary can replay stale~\cite{ReplayStale} or non-optimal~\cite{hao2018end} DNSSEC responses. Replaying stale responses can lead to Pharming when the association between a domain name and an IP address has changed, or can lead to blocking access to a newly added subdomain when NSEC records are replayed.

Regarding availability, DNSSEC is neither \AResDoS{} nor \AANSDoS, since it relies on UDP in both stages and does not provide any anti-DoS mechanisms at the application layer. Thus, adversaries can trigger recursive resolvers and ANSes to expend their resources on spoofed queries. In addition, DNSSEC is not \SECensor{}, because queries are transmitted in plaintext, allowing a censoring agent to block or alter DNS queries/responses in Stage~1. DNSSEC also fails to provide any of the privacy properties, as it was not designed with privacy-related goals. Finally, DNSSEC results in a considerable computation and bandwidth overhead, along with complicated key management with frequent signature generation that should not be overlooked when assessing its provided properties~\cite{cowperthwaite2010futility}.

\subsection{DNSCurve}
\begin{figure}[h!]
\centering
\includegraphics[width=\linewidth]{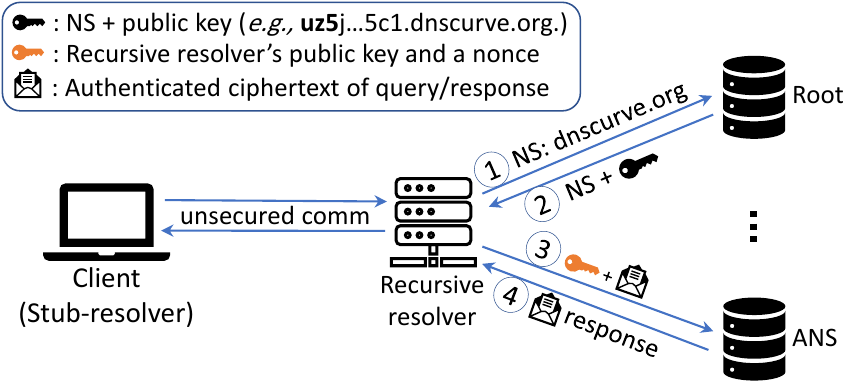}
\caption{\textbf{DNSCurve}: Using authenticated encryption to securely transmit DNS messages between recursive resolvers and ANSes.}
\label{fig:dnscurve}
\end{figure}

As a more secure alternative to DNSSEC, DNSCurve~\cite{DNSCurveweb, curvedraft} uses authenticated encryption based on keys established with Curve25519 Elliptic Curve (EC) to securely transmit DNS messages in Stage~2. The public keys of ANSes are encoded using Base32 and concatenated as subdomains to the domain names of ANSes (\eg “uz5jm...235c1.dnscurve.org”). These concatenated public keys are 54 bytes long and include a hard-coded string `uz5’ at the beginning, which indicates support for DNSCurve by an ANS. A DNSCurve-supporting recursive resolver includes its public key in a DNS query, alongside the encrypted query and a nonce, when it wants to communicate with a DNSCurve-supporting ANS~\cite{curvedraft}. As illustrated in Figure~\ref{fig:dnscurve}, Stage~1 remains unsecured. Unlike DNSSEC, which uses the AD bit to inform clients regarding verification of signatures, DNSCurve does not inform clients about the use of DNSCurve in Stage~2. Clients must therefore blindly trust the recursive resolver and ANSes to correctly implement and use DNSCurve~\cite{dnssecvs}.

In Stage~2, when the recursive resolver queries the name servers (\ie NS records) of `dnscurve.org' from its parent zone (org.) name server (Step 1), the TLD (org.) name server returns the NS records of `dnscurve.org' (Step 2). The NS records of `dnscurve.org' begin with the encoded public key indicated by `uz5' hard-coded string (\eg `uz5j…t35c1' in Figure~\ref{fig:dnscurve}). The recursive resolver then uses the name server's public key and its own private key to generate a shared secret, which is combined with a unique nonce to encrypt a DNS query and create a Message Authentication Code (MAC)~\cite{curvedraft,bernstein2009cryptography}. Then the resolver sends the encrypted query with its public key and nonce to the `dnscurve.org' name server (Step 3). The name server generates the same keystream using its private key, the resolver's public key, and the nonce, then decrypts and verifies the query, generates an authenticated and encrypted response, and sends it back to the recursive resolver (Step 4).

The Salsa20 stream cipher is used to encrypt DNS messages; therefore, DNSCurve is \PEavesTwo{}. Since DNS messages are encrypted, if ECS is included in queries, the intermediate entities in Stage~2 cannot access the included ECS. However, ANSes still have access to the ECS, and DNSCurve partially provides \PHidesTwo{}. Poly1305 MAC is used with the first 32 bytes of the keystream to provide message authentication and thus ensure \SANSImp{}. Due to the use of different established keys for each resolver-ANS communication, a response sent to one resolver cannot be replayed to another~\cite{DNSCurveweb}. In addition, the inclusion of a per-message nonce adds uniqueness to DNS messages, ensuring that one DNS response cannot be replayed to the same resolver in later queries. As a result, DNSCurve provides \SANSReplay{}. However, because long-term private keys must be present on ANS servers to enable real-time authenticated encryption, these keys are exposed to attacks targeting the servers. For this reason, DNSCurve does not satisfy the \SKeyExp{} property.

\subsection{ss2DNS}
\begin{figure}[h!]
\centering
\includegraphics[width=\linewidth]{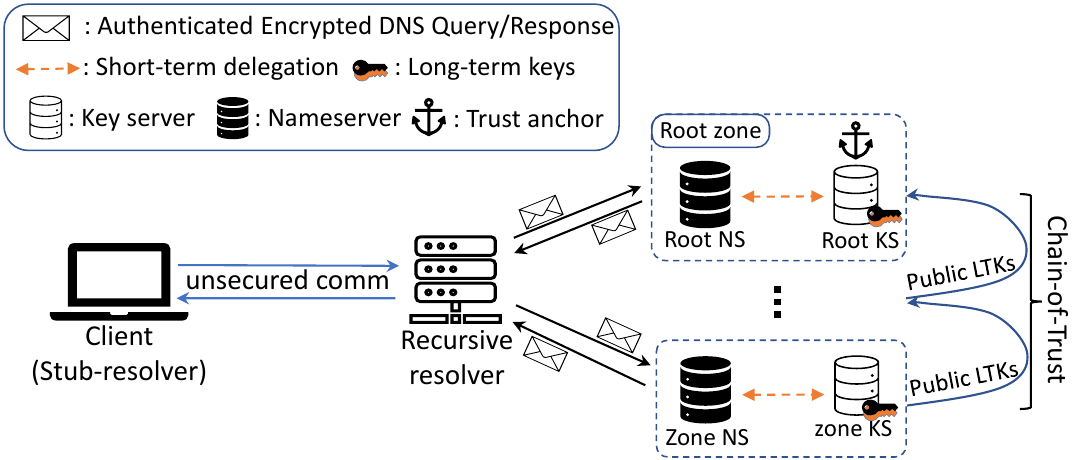}
\caption{\textbf{ss2DNS}: Transferring encrypted and authenticated DNS messages and using a short-term delegation to mitigate the duplication of long-term secrets.}
\label{fig:ss2dns}
\end{figure}
As a secure DNS scheme, ss2DNS~\cite{jahromi2024DNS} is designed to provide security and privacy in Stage~2 of the DNS resolution process. It ensures the confidentiality and integrity of DNS responses through authenticated encryption. In addition, it provides an optional mechanism for transmitting authenticated and encrypted queries within a single round-trip name resolution, and its security and privacy properties supported by a formal analysis~\cite{FormalAnalysis}.

To protect the long-term keys of a zone, ss2DNS introduces a short-term delegation mechanism. This eliminates the need to replicate the zone's long-term key on its nameserver instances, which may not be fully trusted by the zone administrator. By doing so, ss2DNS mitigates the risk of key exposure to potential attacks targeting the nameserver instances. In ss2DNS, separate cryptographic keys are used for securing queries and responses. As illustrated in Figure~\ref{fig:ss2dns}, each zone in ss2DNS has a key server responsible for generating, storing, and managing the zone's long-term keys. Two distinct long-term keys are maintained: one enables secure query transmissions, and the other supports short-term delegation to the zone’s nameserver instances, which then use this delegation to securely transmit DNS responses.

The trust model of ss2DNS follows the same hierarchical structure as DNSSEC. The public portions of both long-term keys are transmitted to the parent zone, where they are published and made available during resolution. This enables the construction of a verifiable chain of trust extending to the root zone. The root zone’s public keys serve as trust anchors and are included in resolver software.

The long-term query key is an agreement key, allowing resolvers to derive encryption keys for securely transmitting queries using authenticated encryption. The second long-term key is used by the zone's key server to sign cryptographic keys of nameserver instances during the delegation phase, authorizing them for a short period to generate authenticated and encrypted responses.

ss2DNS satisfies \SANSImp{}, \PEavesTwo{}, and \PHidesTwo{} by using authenticated encryption for transmitting DNS messages in Stage~2. It also provides \SKeyExp{} by employing short-term delegation within zones and securely storing long-term secrets on a key server. The use of ephemeral keys to derive the response authentication and encryption keys prevents the replay of previous responses to the same or different clients, thereby achieving \SANSReplay{}. However, similar to the other Stage~2 schemes that prioritize performance, the use of UDP without source address validation leaves ANSes susceptible to DoS attacks; therefore, ss2DNS does not satisfy \AANSDoS.

\subsection{DNSCrypt*V3}
\begin{figure}[h!]
\centering
\includegraphics[width=\linewidth]{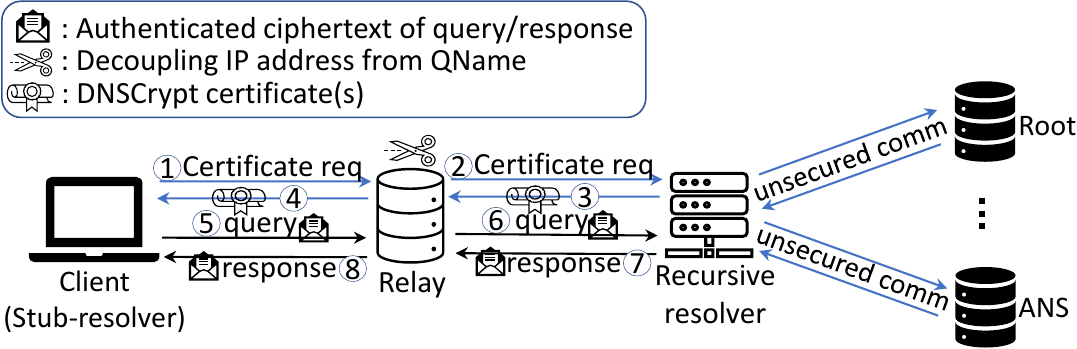}
\caption{\textbf{DNSCrypt*V3}: Provides security properties in Stage~1, and includes a relay server to enhance the privacy and anonymity of clients.}
\label{fig:dnscrypt}
\end{figure}
DNSCrypt~\cite{DNSCryptV2, DNSCryptV3} uses ECC to establish session keys and to transfer authenticated ciphertext of queries and responses at the application layer. Regular DNS queries and responses in Stage~1 are transmitted as authenticated and encrypted messages through software implemented on clients (\eg dnscrypt-proxy) and recursive resolvers (\eg dnscrypt-wrapper). As shown in Figure~\ref{fig:dnscrypt}, in the first step, the client queries a DNSCrypt resolver for its certificates. DNSCrypt does not rely on the Internet's PKI, instead, the client must know the resolver's name and signing public key in advance to verify the certificates and authenticate the resolver~\cite{DNSCryptV2}. The public key of a recursive resolver and its other properties, such as IP address, port number, and provider name, are encoded in the form of \emph{DNS stamps} and embedded in the client-side software. DNSCrypt certificates are encoded as TXT records and follow a specific structure as described in its documentation~\cite{DNSCryptV2}. 

In Step~2, the resolver returns one or more signed certificates, and the client verifies them using a previously distributed public key of the resolver. The client selects the valid certificate with the highest serial number among the supported ones. Using the public key and other cryptographic parameters from the selected certificate, the client generates an authenticated and encrypted query based on the NaCl cryptographic library~\cite{bernstein2009cryptography} and sends the query alongside its own public key to the resolver (Step 3). The resolver decrypts the received query and then sends back an encrypted and authenticated response to the client (Step 4). 

In DNSCrypt version 3, sending anonymized queries were introduced to enhance client privacy. As shown in Figure~\ref{fig:dnscrypt}, DNSCrypt uses an intermediate relay server between the client and the recursive resolver to hide the IP address of the client from the resolver~\cite{DNSCryptV3}. The relay server does not have access to the queried domain name and the recursive resolver does not have access to the client's IP address. This ensures privacy and anonymity with low overhead~\cite{DNSCryptV3}. In practice, a list of public relay servers is maintained by community volunteers.\footnote{\url{https://github.com/DNSCrypt/dnscrypt-resolvers/blob/master/v3/relays.md}} If the relay owner is the same as the owner of the resolver (or if they collude), this one-layer anonymization technique is compromised~\cite{schmitt2019oblivious}. Furthermore, if ECS is included in the client’s DNS queries, the anonymization provided by this method becomes ineffective.

The key exchange in DNSCrypt is performed using the Curve25519 elliptic curve and the hSalsa20 hash function~\cite{DNSCryptV2}. The Salsa20 or ChaCha20 stream cipher is used with Poly1305 for authenticated encryption, providing confidentiality, integrity, and authenticity of transferred messages~\cite{DNSCryptV2}. Additionally, DNSCrypt requires a 24-byte nonce as part of the authenticated ciphertexts, which provides anti-replay protection by adding uniqueness to DNS messages. To enhance privacy by obfuscating packet sizes, DNSCrypt uses the ISO/IEC 7816-4 format to pad DNS messages to a multiple of 64 bytes~\cite{DNSCryptV2}. DNSCrypt runs over port 443, using both TCP and UDP. However, due to its distinguishable characteristics, DNSCrypt traffic is not completely merged with regular web traffic~\cite{dnscryptIsolation}. As a result, DNSCrypt traffic can be isolated through analysis of its distinguishable characteristics~\cite{dnscryptIsolation}.

By employing the Poly1305 message authentication code, DNSCrypt provides \SResImp. Using different keys per client-resolver communication prevents replay of messages among clients, and the nonces included in each message mitigate replay of old messages to the same client. Thus, DNSCrypt satisfies \SResReplay{}. Similar to other Stage~1 schemes, we assume that a simple padding scheme combined with encryption cannot resist traffic analysis attacks, and an adversary is able to infer the visited websites by clients from the encrypted DNS traffic (\eg using ML-based techniques~\cite{siby2019encrypted}). Hence, in our evaluation, DNSCrypt receives partial credit for \PEavesOne{}. With additional effort, censoring entities in Stage~1 can infer the website being visited from the encrypted DNS traffic and impose censorship. Thus, DNSCrypt receives partial credit for \SECensor{}. The default port number for DNSCrypt is 443. Blocking only DNS servers on port 443 is not feasible without disrupting HTTPS connections. However, isolating and blocking access to DNSCrypt resolvers is possible due to the protocol's traffic characteristics. For this reason, DNSCrypt does not satisfy \PConceal{} or \ServerCensor{}. DNSCrypt also does not satisfy \AResDoS{} because, if a DNSCrypt resolver supports UDP transport protocol (as recommended in the specification), an adversary can send spoofed requests and force the resolver to expend computing resources to decrypt and resolve arbitrary DNS queries. The use of intermediate relay servers between clients and recursive resolvers provides a mechanism to hide client IP addresses from resolvers. However, since the DNSCrypt specification does not explicitly mention removal or truncation of ECS, it does not provide \PDecouple.\footnote{Recall that in \hyperlink{P3}{P3}, the assumption is that ECS is included in DNS queries.}

\subsection{Strict DNS-over-TLS (DoT)}
\label{sec:dotsub}
\begin{figure}[h!]
\centering
\includegraphics[width=\linewidth]{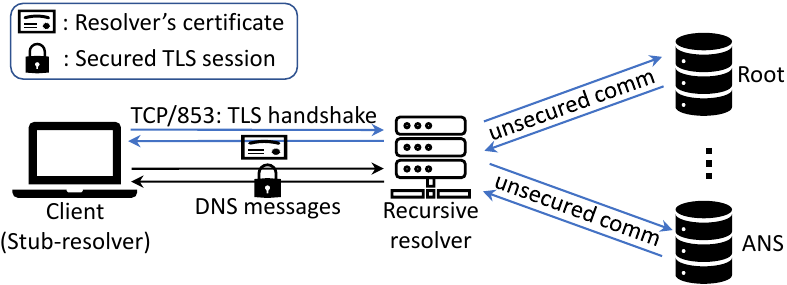}
\caption{\textbf{DoT}: Sending DNS queries over TLS in Stage~1.}
\label{fig:dot}
\end{figure}

DoT~\cite{DoT-rfc7858, zhu2015TDNS} was proposed to securely transfer DNS messages in Stage~1 using Transport Layer Security (TLS). Figure~\ref{fig:dot} illustrates the DoT operation procedure. A client initiates a TLS handshake with a DoT resolver and verifies the resolver’s certificate using PKI trust anchors or an out-of-band channel. The trust anchors may vary depending on stub resolver configurations or the client’s operating system~\cite{rootstoresAbdou}. Similar to HTTPS, DoT uses X.509 certificates to bind the public key of resolvers to their owner. Relying on the PKI means that DoT inherits any vulnerabilities (\eg compromised CAs) and security weaknesses in any external PKI on which it depends. After the TLS handshake is completed, the client and the resolver establish a secure session with a shared secret to transfer DNS messages~\cite{DoT-rfc7858}. DoT supports two privacy modes: \emph{strict} and \emph{opportunistic}. In the \emph{opportunistic} mode, encryption and authentication are used when available and successfully negotiated~\cite{DoT-rfc7858}. Otherwise, the connection falls back to Vanilla DNS. In the \emph{strict} mode, both successful authentication and encryption are mandatory for resolving domain names. In the presence of an active attacker, the \emph{opportunistic} mode may mislead clients regarding the level of security provided, as the connection can be downgraded to insecure Vanilla DNS.

When DoT is used with persistent TLS connections, it introduces negligible overhead compared to Vanilla DNS~\cite{lu2019end}. However, maintaining the excessive number of open connections can exhaust the resolver resources. Regarding anonymity, DoT does not provide any mechanism to hide the client's IP address from the resolvers. \SResImp{} is achieved by TLS-based authenticated messages. Freshly generated session keys for each handshake prevent replay from one session to others, while per-record nonces prevent replay of messages within the same session. Together, these mechanisms enable \SResReplay{}. DoT suggests using the EDNS(0) padding option to pad the DNS messages with a variable number of octets (\eg filled with 0x00), thereby improving resilience against traffic analysis and side-channel leaks in Stage~1. However, other studies demonstrate that the encryption and padding in DoT are insufficient to resist ML-based traffic analysis attacks~\cite{siby2019encrypted, DoTLeak}. As a result of traffic analysis, the visited websites by users will be inferred, and name resolution can be censored based on that information. Thus, DoT receives partial credit for \PEavesOne{} and \SECensor{} since traffic analysis incurs additional overhead and effort for the censoring/eavesdropping entities, and the accuracy and efficacy of this approach vary based on the trained models and selected traffic features. TCP ensures that spoofed queries cannot be processed processed at the application layer without completing the handshake at the transport layer. In addition, several mitigation measures have been proposed to improve the resistance of TCP to SYN flooding attacks~\cite{rfc4987}. Therefore, DoT provides \AResDoS{}. However, DoT does not provide \ServerCensor{} or \PConceal{}, because it uses the dedicated port number 853. As a result, its traffic is easily distinguishable, and thus the DoT servers susceptible to censorship.

\subsection{DNS-over-HTTPS (DoH)}
\label{sec:dohsub}
\begin{figure}[h!]
\centering
\includegraphics[width=\linewidth]{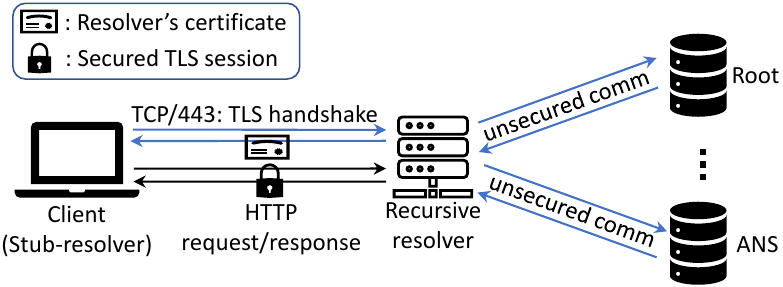}
\caption{\textbf{DoH}: Sending DNS queries over HTTPS using URIs and HTTP methods}
\label{fig:doh}

\end{figure}

DoH~\cite{DoH-rfc8484} encodes DNS messages as HTTP requests and responses, sending them securely over TLS using the format specified by HTTP Uniform Resource Identifier (URI) templates (\eg https://dns.server.com/dns-query?). Since DoH operates on top of TLS, it provides many of the same properties as strict DoT in Stage~1. Resolver domain names are associated with their public keys through X.509 certificates issued by CAs. Therefore, similar to DoT, DoH inherits the limitations and vulnerabilities of any external PKI upon which it depends. Unlike DoT, DoH does not support the opportunistic mode, meaning it does not fail open in case of invalid certificates. If certificate validation fails, name resolution fails. As a result of this strict security in DoH, Lu \etal~\cite{lu2019end} did not find DoH resolvers with invalid certificates in their measurement. Furthermore, DoH uses port 443, the same port as HTTPS, thereby merging DNS traffic with web traffic and making its traffic less distinguishable~\cite{lu2019end, DoH-rfc8484}.  However, traffic analysis techniques can still be used to isolate DoH traffic from other HTTPS traffic, although doing so requires additional effort, and the accuracy of such isolation depends on factors such as trained machine learning models and selected traffic features~\cite{vekshin2020doh}. Thus, DoH receives half credit for \PConceal{}. Consequently, blocking access to DoH servers based on isolated traffic is feasible, and DoH partially provides \ServerCensor{}.

DoH also supports HTTP/2 compression, in addition to EDNS(0) DNS padding, to enhance the privacy of transmitted messages and improve resilience against traffic analysis and side-channel leaks~\cite{DoH-rfc8484, EDNS-Padding}. However, similar to DoT, Siby \etal~\cite{siby2019encrypted} demonstrated that simple padding schemes are insufficient to prevent ML-based traffic analysis attacks that reveal the pages visited by users, although they increase the cost of such analysis for adversaries. Since DoH opeartes on top of TLS, its remaining properties are similar to those of DoT.

\subsection{Strict DNS-over-QUIC (DoQ)}
\begin{figure}[h!]
\centering
\includegraphics[width=\linewidth]{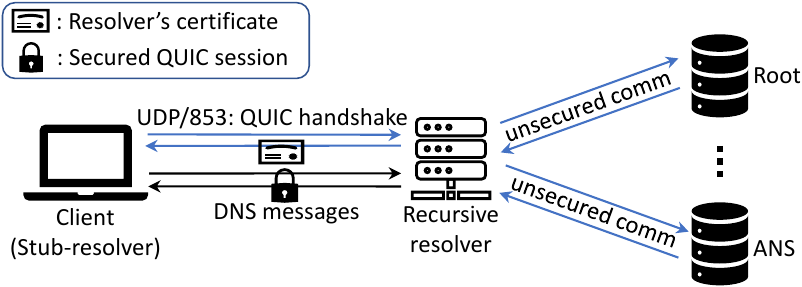}
\caption{\textbf{DoQ}: Sending DNS using QUIC transport protocol.}
\label{fig:DoQ}
\end{figure}

DoQ~\cite{rfc9250} uses QUIC~\cite{rfc9000, rfc9001} transport-layer protocol as a general-purpose scheme that can improve the name resolution process and zone transfer.\footnote{Herein, we analyze DoQ in Stage~1.} As illustrated in Figure~\ref{fig:DoQ}, the DoQ workflow is similar to DoT, but it achieves a faster handshake in DoQ by combining the TCP and TLS handshakes into a single QUIC handshake. DoQ improves the performance of name resolution and solves some known problems in TCP (\eg head-of-line blocking~\cite{bottger2019empirical}) and UDP (\eg IP fragmentation~\cite{herzberg2013fragmentation}). In 2022, over 1200 DoQ resolvers were discovered on the Internet~\cite{kosek2022one}. By default DoQ uses the designated port number UDP/853; thus, its traffic is distinguishable and access to DoQ servers can be blocked. However, the specification also recommends the use of port UDP/443, which merges DoQ traffic with HTTP/3 and thereby reduces the effectiveness of port-based blocking~\cite{rfc9250}. DoQ provides security properties, usage profiles, and resolver authentication methods similar to those of DoT~\cite{rfc9250, rfc8310}. Similar to DoT, we evaluate DoQ under the \emph{strict} usage profile.

The specification recommends using EDNS(0) padding or QUIC packet padding in implementations to mitigate traffic analysis attacks~\cite{rfc9250}. Padding at the QUIC packet level is preferred over EDNS(0) padding, as it provides better performance while providing the same level of protection~\cite{rfc9250}. To the best of our knowledge, no study has yet demonstrated traffic analysis attacks that reveal the visited websites by users in DoQ. However, for web traffic, QUIC padding does not eliminate fingerprinting attacks, and other application-layer defenses (\eg injecting cover traffic) have been shown to be more effective~\cite{QUIC-WF}. Thus, similar to DoT and DoH, we assume that with simple padding strategies DoQ remains susceptible to traffic analysis attacks that reveal the visited web pages and it receives partial credit for \PEavesOne{}. Furthermore, because DoQ uses a designated port number, its traffic is easily distinguishable and it does not receive credit for \PConceal{} and \ServerCensor{}. The remaining DoQ properties are the same as those of DoT.

\subsection{DNS-over-DTLS (DoDTLS)}
\begin{figure}[h!]
\centering
\includegraphics[width=\linewidth]{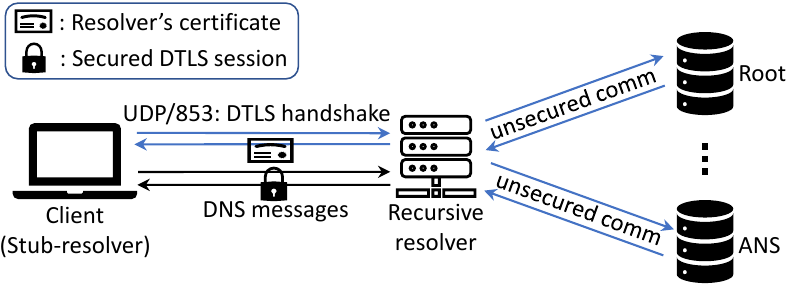}
\caption{\textbf{DoDTLS}: Sending DNS using DTLS transport protocol.}
\label{fig:dodtls}
\end{figure}

DoDTLS~\cite{dodtls8094} uses Datagram Transport Layer Security (DTLS) to secure Stage~1 DNS messages over port number UDP/853. DoDTLS addresses head-of-line blocking and provides improved performance compared to DoT. DTLS has security properties comparable to TLS, with one of the main differences being the handshake, which contains additional header fields, cookies, and retransmission timers for error handling and DoS attack prevention~\cite{rfc9147}. DoDTLS authentication mechanisms are similar to those of DoT; therefore, it inherits the same PKI vulnerabilities and complexities that affect DoT.

Similar to DoT, EDNS(0) padding could be implemented in DoDTLS. However, encryption and simple padding schemes do not significantly improve the privacy of DNS messages in TLS-based alternatives~\cite{siby2019encrypted}. To the best of our knowledge, there are no real-world implementations of DoDTLS. Here, we assume that the padding scheme similar to DoT is not resilient to traffic analysis attacks, and DoDTLS receives partial credit for \PEavesOne{}. In the strict DoDTLS privacy profile, a client must complete the DTLS handshake with the recursive resolver before sending a DNS query. Consequently, the query is not processed at the application layer on the recursive resolver prior to the completion of the handshake. This approach makes DoDTLS resistant to resolver DoS attacks and provides \AResDoS{}. The remaining properties of DoDTLS are the same as DoT.

\subsection{DNS-over-Tor (DoTor)}
\begin{figure}[h!]
\centering
\includegraphics[width=\linewidth]{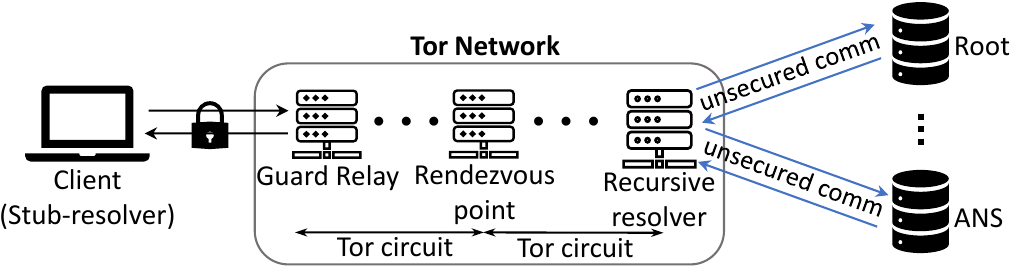}
\caption{\textbf{DNS-over-Tor}: Based on Cloudflare's hidden resolver structure~\cite{dotor}.}
\label{fig:dotor}
\end{figure}

DoTor can be implemented in different ways, such as a hidden resolver or a regular public resolver over the Internet queried through the Tor network. In April 2018, Cloudflare implemented a hidden resolver for DoTor~\cite{dotor}.\footnote{Cloudflare's DoTor resolver supports DoT, DoH, and Vanilla-DNS queries.} Tor is vulnerable to several types of attacks~\cite{DNSonTor, ExitRelays, Badrelays, TorCorrelation}. One example is traffic correlation, where incoming and exit traffic of the Tor network can be analyzed to deanonymize clients~\cite{TorCorrelation, DNSonTor}. Malicious relays~\cite{Badrelays} pose another threat, as they can perform different types of manipulation, redirection, or eavesdropping to compromise anonymity or manipulate clients' traffic~\cite{Badrelays, ExitRelays}. DoTor can also be deployed as a `.onion' service to mitigate specific Tor correlation attacks, such as the DefecTor attack, where the adversaries analyze exit relay name resolution to perform correlation attacks~\cite{dotor, DNSonTor}. By using a hidden `.onion' recursive resolver, no name resolution occurs at exit nodes; instead, resolution is performed using distributed hash table lookups within the Tor network. This mitigates DefecTor attacks that could otherwise reveal the recursive resolver used by a client~\cite{DNSonTor}. Figure~\ref{fig:dotor} illustrates the DoTor working procedure. First, the client establishes a secure session with the hidden resolver through the onion routers, including a rendezvous point~\cite{dingledine2004tor}. TLS is then used to establish authenticated, encrypted sessions between the Tor nodes. The DNS traffic is encrypted using the established keys and is sent through the Tor network to the hidden resolver. The entry node to the Tor network that the client directly communicates with is called the guard relay. Upon receiving a query, the hidden resolver traverses the DNS hierarchy insecurely and sends back the response to the client through the secured session.

DoTor ensures the confidentiality of DNS messages by applying layers of encryption on fixed-sized units of data, which preserves clients' privacy and provides a high degree of anonymity from intermediate entities~\cite{dingledine2004tor}. As a result, ISPs cannot identify DNS traffic or correlate it with other traffic in order to reveal requested domain names. Moreover, Siby \etal~\cite{siby2019encrypted} demonstrated that DoH-over-Tor is one of the secure DNS schemes that is significantly resilient against ML-based traffic analysis attacks that infer visited web pages based on encrypted DNS traffic. However, the additional layers of encryption and TLS communication between relays can significantly increase the time and size overhead compared to other alternatives, since Tor inherently sacrifices latency for security and anonymity. Depending on the user or application, the security benefits of sending DNS messages over Tor may outweigh the overhead caused by DoTor~\cite{TorMuffett}. Hidden resolvers also have some degree of protection against DoS attacks as they are placed behind the Tor network and benefit from its distributed and load-balanced architecture. Thus, DoTor hidden resolvers are resilient to DoS attacks. Furthermore, since Tor only supports TCP for internal communication, Tor nodes can implement SYN cookies and rate limiting to mitigate DoS attacks.\footnote{\url{https://support.torproject.org/abuse/}}

DoTor is the only protocol among the evaluated schemes that provides complete \PConceal{} by encapsulating messages into fixed-size Tor cells, making DNS traffic indistinguishable~\cite{siby2019encrypted}. Furthermore, no specific port number is assigned to Tor communication, so port 443 can be used to combine Tor with web traffic. The intermediate Tor nodes decouple a client’s IP address from the recursive resolver, thereby providing \PDecouple{}. Additionally, the use of short-term ephemeral keys in Tor circuit communications mitigates replay attacks between different sessions. TLS per-record nonces can mitigate replay of messages in the same session, meaning DoTor provides \SResReplay{}. Although censoring governments cannot directly censor the recursive resolvers in DoTor, they can restrict access to the resolvers by blocking the entire Tor-related traffic (\eg through deep packet inspection of Tor traffic or by blocking access to Tor-related nodes)~\cite{afroz2021timeline}. Thus, DoTor does not provide \ServerCensor. In Cloudflare’s implementation, all included ECS data is removed from DNS queries [86]; therefore, DoTor satisfies \PHidesTwo.

\subsection{Oblivious DNS (ODNS)}
\begin{figure}[h!]
\centering
\includegraphics[width=\linewidth]{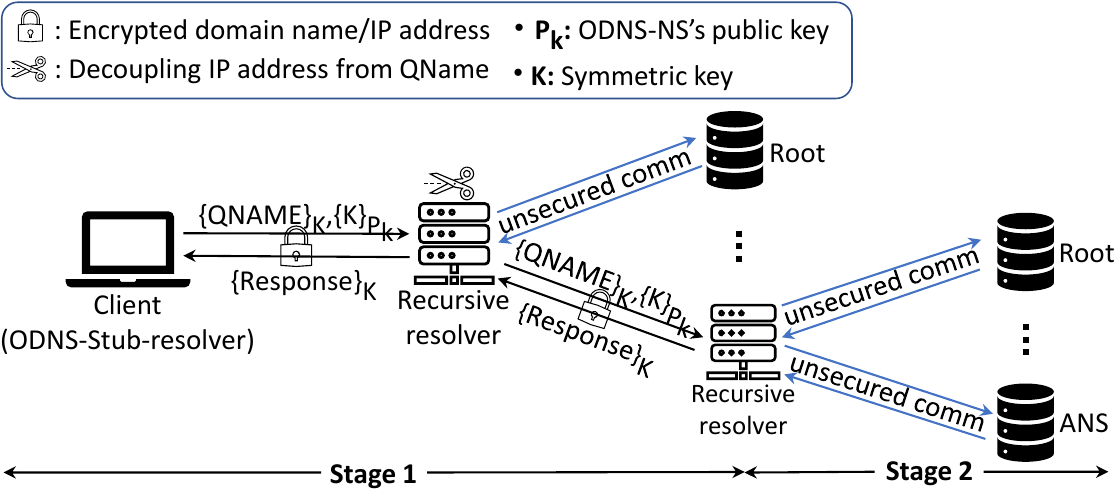}
\caption{\textbf{Oblivious DNS}: Adding confidentiality and anonymity to DNS queries~\cite{schmitt2019oblivious}.}
\label{fig:odns}

\end{figure}

Schmitt et al.~\cite{schmitt2019oblivious} proposed ODNS to preserve client privacy and provide anonymity from recursive resolvers by decoupling a client's IP address from the queried domain name and encrypting the domain name, resulting in a negligible page load time overhead. An ODNS name server does not have access to the client's IP address, while the recursive resolver, which acts as a relay in Stage~1, does not have access to the client's queried domain name. Before name resolution begins, the client must obtain the public key of the ODNS-NS $P_k$. To do so, the client sends a special query to the recursive resolver, which uses an anycast address to return the public key of the closest ODNS-NS, included in the EDNS extension of the response message. Initially, the client obtains the public key of the ODNS-NS. Then, as illustrated in Figure~\ref{fig:odns}, the client encrypts the domain name in the DNS query with a newly generated symmetric key $K$, and then encrypts $K$ with $P_k$~\cite{schmitt2019oblivious}. The query and the encrypted symmetric key `$\{K\}_{P_k}$' are sent to the recursive resolver, which passes the DNS query to the ODNS-NS using its IP address as the source address. The ODNS-NS then traverses the DNS hierarchy to resolve the queried domain name, encrypts the final response with the client's symmetric key $K$, and sends it back to the client~\cite{schmitt2019oblivious}.

In Stage~1, the middle entities between the client and the recursive resolver know the client's IP but not the queried domain name, and the entities between the recursive resolver and the ODNS-NS neither know the queried domain name nor the client's IP address. ODNS removes ECS from DNS queries in the stub resolvers to preserve clients' privacy~\cite{schmitt2019oblivious}. Therefore, even if the client adds its IP address to the queries in the ECS, an ideal ODNS stub resolver forwards the query without it. Therefore, ODNS completely provides \PHidesTwo{}. ODNS does not provide any other property in Stage~2 as it was proposed for Stage~1.

ODNS provides confidentiality and anonymity against passive adversaries that do not have access to the messages on the entire DNS resolution path. Similar to DoT and DoH, encryption in ODNS without proper padding or traffic-flow security techniques makes the encrypted DNS traffic susceptible to traffic analysis attacks that can reveal the visited web pages. Furthermore, the entire payload of DNS queries and responses is not encrypted, and sections such as ``Additional" or ``Authority" in DNS responses can disclose information about the queried domain name. Thus, ODNS receives partial credit for \PEavesOne{} and \SECensor{}. This scheme also provides \PDecouple{} using a one-layer proxy, and the ECS is removed at the stub resolvers. Since the integrity and availability of DNS requests are not among the design goals of ODNS, it does not provide properties related to these goals.

\subsection{Confidential DNS and IPsec-based DNS Security}
\begin{figure}[h!]
\centering
\includegraphics[width=\linewidth]{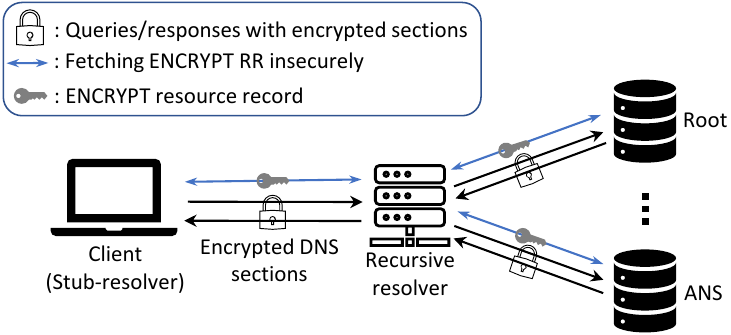}
\caption{\textbf{Confidential-DNS}: Adding confidentiality in both stages of DNS.}
\label{fig:confdns}
\end{figure}

Another scheme proposed as an Internet-Draft, but which did not progress into an RFC, was Confidential DNS~\cite{confdns}. Confidential DNS introduces a new resource record, to DNS known as \texttt{ENCRYPT}, which contains the public key of a recursive resolver or ANS. Clients use the public key obtained from the \texttt{ENCRYPT} RR to securely transfer a shared secret or public key, which is used to encrypt DNS fields in queries and responses~\cite{confdns}.
Confidential DNS works in two modes: \emph{opportunistic} and \emph{authenticated}. In the former, the client does not verify the keys received from a DNS server. In the latter, authentication and verification of the Confidential DNS keys occur through DNSSEC~\cite{confdns}. In Confidential DNS, the authenticity of DNS messages, which DNSSEC has provided, is not among the goals of Confidential DNS. As a result, Confidential DNS often provides Opportunistic Encryption (OE) (\ie encryption is used when possible, otherwise it falls back to plaintext) without authentication.

A similar approach to Confidential DNS is the use of IPsec-based OE to secure the transmission of DNS messages~\cite{IPSECA}. In this scheme, a new resource record, \texttt{IPSECA}, is added to recursive resolvers or ANSes to enable IPsec tunnels and secure transmission of DNS messages using IPsec-related security properties at the network layer~\cite{IPSECA}. DNSSEC signatures and chain of trust must be used to verify \texttt{IPSECA} records. After setting up IPsec Security Policy Database (SPD), the \texttt{IPSECA} recipient verifies the Security Association (SA) credentials with the received IPSECA record~\cite{IPSECA}. In this work, we evaluate Confidential DNS and IPsec-based OE without DNSSEC-based authentication. Active network-based adversaries can replace the transmitted key records with false ones, thereby misleading clients regarding the protection in place. Thus, based on our threat model, no property is provided by opportunistic Confidential DNS or IPsec-based OE, and these schemes are not evaluated in Table~\ref{tab:comptable}.
\begin{table*}[ht!]
\begin{minipage}{\textwidth}
\caption[Comparative evaluation framework for secure DNS schemes]{Comparative evaluation framework for assessing properties of secure DNS schemes (s1: Stage~1, s2: Stage~2)}
\label{tab:comptable}
\raggedright
\resizebox{0.9\textwidth}{!}{
    \begin{tabular}{@{}c@{}|@{}c@{}|@{}c@{}|RRGGGBBB|RRRGBB}
        & & &  \multicolumn{8}{c}{\textbf{Stage~1}} & \multicolumn{6}{c}{\textbf{\begin{tabular}[c]{@{}c@{}}Stage~2\end{tabular}}}  \\
          \multicolumn{1}{l|}{\adjustbox{angle=0,lap=\width-0.5em}{\textit{DNS Scheme}}} &
           \multicolumn{1}{c|}{\adjustbox{angle=90,lap=\width-0.5em}{\textit{Reference}}} &
            \multicolumn{1}{c|}{\adjustbox{angle=90,lap=\width-0.5em}{\textit{Proposed Working Stage}}} &
              \headrow{\textit{\hyperlink{S1}{S1.~\SResImp(s1)}}} &
               \headrow{\hyperlink{S2}{\textit{S2.~\SResReplay(s1)}}}  &
                \headrow{\hyperlink{A1}{\textit{A1.~\SECensor}}} &
                 \headrow{\hyperlink{A2}{\textit{A2.~\ServerCensor}}} &
                  \headrow{\hyperlink{A3}{\textit{A3.~\AResDoS}}} &
                   \headrow{\hyperlink{P1}{\textit{P1.~\PEavesOne}}} &
                    \headrow{\hyperlink{P2}{\textit{P2.~\PConceal}}} &
                     \headrow{\hyperlink{P3}{\textit{P3.~\PDecouple}}} &
                      \headrow{\hyperlink{S3}{\textit{S3.~\SANSImp(s2)}}} &
                       \headrow{\hyperlink{S4}{\textit{S4.~\SANSReplay(s2)}}} &
                       \headrow{\hyperlink{S5}{\textit{S5.~\SKeyExp}}} &
                        \headrow{\hyperlink{A4}{\textit{A4.~\AANSDoS}}} &
                         \headrow{\hyperlink{P4}{\textit{P4.~\PEavesTwo}}} &
                          \headrow{\hyperlink{P5}{\textit{P5.~\PHidesTwo}}} \\ \hline

\multicolumn{1}{l|}{Vanilla DNS} & \cite{rfc1034,rfc1035} & \multirow{2}{*}{1\&2} & \Circle & \Circle & \Circle & \Circle & \Circle & \Circle & \Circle & \Circle & \Circle & \Circle & \Circle & \Circle & \Circle & \Circle  \\ 

\multicolumn{1}{l|}{S-DNS} & \cite{bassil2012security} &  & \CIRCLE & \CIRCLE & \Circle & \Circle & \Circle & \Circle & \Circle & \Circle & \CIRCLE & \CIRCLE & \Circle & \Circle & \Circle & \Circle \\ \hline

\multicolumn{1}{l|}{DNSSEC} & \cite{rfc4033,rfc4034,rfc4035} & \multirow{4}{*}{2} & \Circle & \Circle & \Circle & \Circle & \Circle & \Circle & \Circle & \Circle & \CIRCLE & \Circle & \CIRCLE & \Circle & \Circle & \Circle  \\ 

\multicolumn{1}{l|}{Live DNSSEC} & \cite{liveDNSSEC} &  & \Circle & \Circle & \Circle & \Circle & \Circle & \Circle & \Circle & \Circle & \CIRCLE & \Circle & \Circle & \Circle & \Circle & \Circle \\ 

\multicolumn{1}{l|}{DNSCurve} & \cite{DNSCurveweb, curvedraft} &  & \Circle & \Circle & \Circle & \Circle & \Circle & \Circle & \Circle & \Circle & \CIRCLE & \CIRCLE & \Circle & \Circle & \CIRCLE & \LEFTcircle \\

\multicolumn{1}{l|}{ss2DNS} & \cite{jahromi2024DNS} &  & \Circle & \Circle & \Circle & \Circle & \Circle & \Circle & \Circle & \Circle & \CIRCLE & \CIRCLE & \CIRCLE & \Circle & \CIRCLE & \LEFTcircle \\ \hline

\multicolumn{1}{l|}{DNSCrypt*V3} & \cite{DNSCryptV3} & \multirow{7}{*}{1} & \CIRCLE & \CIRCLE & \LEFTcircle & \Circle & \Circle & \LEFTcircle & \Circle & \Circle & \Circle & \Circle & \Circle & \Circle & \Circle & \Circle \\ 

\multicolumn{1}{l|}{Strict DNS-over-TLS} & \cite{DoT-rfc7858} &  & \CIRCLE & \CIRCLE & \LEFTcircle & \Circle & \CIRCLE & \LEFTcircle & \Circle & \Circle & \Circle & \Circle & \Circle & \Circle & \Circle & \Circle \\ 

\multicolumn{1}{l|}{DNS-over-HTTPS} &\cite{DoH-rfc8484} &  & \CIRCLE & \CIRCLE & \LEFTcircle & \LEFTcircle & \CIRCLE & \LEFTcircle & \LEFTcircle & \Circle & \Circle & \Circle & \Circle & \Circle & \Circle & \Circle \\ 

\multicolumn{1}{l|}{Strict DNS-over-DTLS} & \cite{dodtls8094} &  & \CIRCLE & \CIRCLE & \LEFTcircle & \Circle & \CIRCLE & \LEFTcircle & \Circle & \Circle & \Circle & \Circle & \Circle & \Circle & \Circle & \Circle \\ 

\multicolumn{1}{l|}{Strict DNS-over-QUIC(S1)} & \cite{rfc9250} &  & \CIRCLE & \CIRCLE  & \LEFTcircle & \Circle & \CIRCLE & \LEFTcircle & \Circle & \Circle & \Circle & \Circle & \Circle & \Circle & \Circle & \Circle \\ 

\multicolumn{1}{l|}{DNS-over-Tor} & \cite{dotor} &  & \CIRCLE & \CIRCLE & \CIRCLE & \Circle & \CIRCLE & \CIRCLE & \CIRCLE & \CIRCLE & \Circle & \Circle & \Circle & \Circle & \Circle & \CIRCLE \\ 

\multicolumn{1}{l|}{Oblivious DNS} & \cite{schmitt2019oblivious} &  & \Circle & \Circle & \LEFTcircle & \Circle & \Circle & \LEFTcircle & \Circle & \CIRCLE & \Circle & \Circle & \Circle & \Circle & \Circle &  \CIRCLE \\ \hline
\end{tabular}
}
\end{minipage}
\end{table*}

\section{Comparative Evaluation}
\label{sec:comparison}
In this section, we compare the surveyed schemes in Section~\ref{sec:alternatives}. The properties are organized in Table~\ref{tab:comptable}, where the columns are the properties defined in Section~\ref{sec:properties} and the rows are the surveyed DNS schemes from Section~\ref{sec:alternatives}. Each cell in Table~\ref{tab:comptable} is marked with a full bullet (\CIRCLE), a half-circle (\LEFTcircle), or an empty circle (\Circle). A bullet in row $i$ column $j$ indicates that the scheme in row $i$ fully satisfies the property in column $j$. A Half-circle means partially satisfies, and an empty circle indicates that the scheme does not satisfy the property. Using this framework, we provide insights regarding associations between properties, conduct a detailed comparison of the schemes, and discuss the overarching challenges of security, privacy, and availability in the DNS resolution process to highlight trends.

\subsection{Association Among Properties}
Before discussing the bigger picture in Table~\ref{tab:comptable}, we clarify the association and difference between properties that have similar ratings. While similarities in ratings my suggest redundancy between properties, they may result from the limited number of schemes evaluated. Whenever applicable, we illustrate such cases by describing a hypothetical DNS scheme that would satisfy one property but not the other.

\textbf{\hyperlink{S1}{S1} and \hyperlink{S2}{S2}:} The properties \SResImp{} and \textit{Resili\-ent-to-Resolver-Replay-Attack} may at first reading appear to be equivalent. \textit{Resilient-to-False-Res\-olver-Response} is provided through implementation of countermeasures against malicious false message injection, such as including MACs or digital signatures in a scheme. \SResReplay{} can be provided by integrating anti-replay mechanisms with a scheme; \eg by using distinct session keys per session to provide session freshness and per-record nonces to ensure freshness of messages in a session. Therefore, it is possible to have a DNS scheme that satisfies \SResImp{} (\eg utilizing MACs) but not \SResReplay{} due to the absence of anti-replay mechanisms. In other words, anti-replayability requires a more robust form of authentication, wherein the interactions between entities have a one-to-one relationship, which inherently implies the simpler form of message authentication without anti-replay mechanisms~\cite{lowe1997hierarchy}.

\textbf{\hyperlink{A1}{A1}, \hyperlink{P1}{P1}, and \hyperlink{P2}{P2}:} If a secure DNS scheme prevents unauthorized access to its messages (\ie fulfills \PEavesOne{} (\hyperlink{P1}{P1})) or renders its traffic indistinguishable (\ie satisfies \PConceal{} (\hyperlink{P2}{P2})), the lack of access to DNS payload prevents censoring entities from DNS payload-based censorship. Therefore, to the extent that a given scheme satisfies either \hyperlink{P1}{P1} or \hyperlink{P2}{P2}, it fulfills \hyperlink{A1}{A1} (\PConceal{}), as \hyperlink{P1}{P1} or \hyperlink{P2}{P2} render plaintext of DNS message inaccessible to adversaries.

\textbf{\hyperlink{P5}{P5} for Stage~1 schemes:} Property \PHidesTwo{} (\hyperlink{P5}{P5}) implies preventing exposure of client IP addresses or subnets in Stage~2. Client IP addresses in Stage~2 are typically included in DNS payload in form of ECS. Removing ECS from the client's query at stub resolvers or recursive resolvers ensures that the client's IP address remains undisclosed in Stage~2. In Stage~1, a scheme that \PDecouple{} (\hyperlink{P3}{P3}) must additionally remove ECS from its DNS queries, as failure to do so would render the added overhead futile. Stage~1 schemes (\ie DoTor or ODNS) that remove ECS from DNS queries at stub or recursive resolvers achieve full credit for \PHidesTwo.

\subsection{Comparison of Schemes}
Several observations can be made from the analysis of Table~\ref{tab:comptable}. First, none of the secure DNS schemes provide security in the whole path of DNS resolution (\ie both stages), and none satisfy all properties in one stage. The majority of secure-DNS schemes in our evaluation are designed to operate in Stage~1. To ensure a secure DNS resolution process, these Stage~1 schemes have to assume that other Stage~2 schemes (\eg DNSSEC) are employed to secure the communication between recursive resolvers and ANSes. However, Stage~2 schemes such as DNSSEC and DNSCurve are as yet not widely deployed and do not provide all of the defined properties in Stage~2. Thus, clients using Stage~1 schemes might falsely assume that their DNS resolution is fully protected, while the security and privacy of their queries can be compromised in Stage~2.

\textbf{DNSSEC (disproportionate effort for minimal return):} Table~\ref{tab:comptable} illustrates that DNSSEC only provides two properties: \SANSImp{} (\hyperlink{S3}{S3}) and \SKeyExp{} (\hyperlink{S5}{S5}). As stub resolvers are typically configured as non-validating and depend on recursive resolvers for DNSSEC usage and validation, DNSSEC only protects Stage~2. Recursive resolvers in Stage~1 use designated bits to signal the clients that a response is DNSSEC-validated. However, the communication between stub resolvers and recursive resolvers remains unprotected. The lack of security in Stage~1 allows an attacker to inject false responses to clients, while possibly misleading them into believing that DNSSEC has provided response validation. DNSSEC is not resilient against DoS attacks, and due to the increased response size is subject to reflection amplification attacks. Furthermore, due to long-lived signatures, DNSSEC-signed resource records are susceptible to replay attacks, enabling the replay of stale records or sub-optimal responses in CDN services~\cite{hao2018end}. Additionally, DNSSEC was not designed to provide privacy-related properties; thus, it does not satisfy any of our privacy properties.

DNSSEC, while providing minimal properties, introduces a significant key management and signature generation burden adding computational and bandwidth overhead to DNS resolution in Stage~2. 

The live DNSSEC implementation in Table~\ref{tab:comptable} denotes a version of DNSSEC that generates signatures in real-time, requiring the presence of signature generation keys on the servers. Since the generated signatures are not unique for each query, this real-time approach remains vulnerable to replay attacks. Furthermore, the requirement to store signature keys on server instances increases the risk of key exposure to potential threats. Consequently, unlike offline-signed DNSSEC records, this implementation does not satisfy \hyperlink{P5}{P5}.

\textbf{Beyond encryption:} All of the evaluated schemes that were proposed to work in Stage~1 use encryption to prevent unauthorized access to plaintext DNS messages. However, encryption alone has been found to be insufficient to protect against traffic analysis attacks. These attacks leverage metadata and encrypted data patterns, such as time, size, direction, or order of packets, to infer or classify the queried domain name or visited websites through statistical or ML-based techniques. Simple padding schemes have been demonstrated to be ineffective in mitigating traffic analysis attacks that reveal the domain names visited by users~\cite{siby2019encrypted}.

One technique for obscuring size-related metadata and patterns is employing large padding schemes prior to encryption. However, this results in a significant increase in bandwidth overhead. An alternative approach is to adopt schemes that repackage traffic into uniform-sized packets, such as Tor cells in DoTor. However, using Tor also results in significant time, computational, and bandwidth overhead. A viable approach for mitigating traffic analysis, which requires further research, can be using various techniques such as padding, merging traffic with other protocols, and implementing traffic-flow security measures (\eg sending cover traffic) in the DNS context to effectively mitigate such attacks with a reasonable overhead.

\textbf{TLS-based DNS schemes:} Among the TLS-based schemes, DoH provides more privacy and availability properties than the others. In addition to the properties of other schemes, DoH partially provides \PConceal{} (\hyperlink{P2}{P2}), and \ServerCensor{} (\hyperlink{A2}{A2}) by merging its traffic with the web. TLS-based schemes that use UDP as the transport layer protocol, such as DoDTLS and DoQ, are designed to enhance the performance of DoT while maintaining similar security and privacy properties.

\textbf{Near-identical schemes:} Except for \PConceal{} (\hyperlink{P2}{P2}) and \textit{Resi\-lient-to-DNS-Server-Censorship} (\hyperlink{A2}{A2}), DNSCrypt*V3 and DoH provide similar properties. Due to the distinguishability of DNSCrypt's traffic, previous research demonstrated that isolating DNSCrypt traffic is relatively easy without sophisticated techniques~\cite{dnscryptIsolation}. However, isolating DoH traffic from web traffic requires more complex ML-based and statistical techniques. Consequently, isolation of DoH traffic is more complex than isolation of DNSCrypt traffic; thus, DoH partially satisfies \PConceal{} and \ServerCensor{}, while DNSCrypt does not. Moreover, DoH does not provide any mechanism to conceal the IP addresses of clients from recursive resolvers; therefore, DoH does not satisfy \PDecouple{} (\hyperlink{P3}{P3}). Although DNSCrypt employs relays to decouple DNS queries from client IP addresses, its lack of explicit removal or truncation of ECS in DNSCrypt rules out credit for \PDecouple{}.

\textbf{Confidentiality in Stage~2:} All of the secure DNS schemes that work in Stage~1 provide confidentiality in that stage. Hence, based on their use of encryption, they all provide some degree of \PEavesOne{} (\hyperlink{P1}{P1}). However, in Stage~2, only DNSCurve provides confidentiality, which has not been adopted by the root and TLDs. That is, the majority of DNS traffic in Stage~2 consists of DNSSEC and Vanilla DNS, with their plaintext data being accessible to unauthorized entities. If Stage~2 traffic contains client-related information, such as ECS or complete QName, it will disclose client-related information in Stage~2. This highlights the importance of eliminating or truncating ECS from DNS messages and employing QName minimization to reduce client-related information leaks in Stage~2.

\textbf{Censorship resilience:} Schemes that use a distinguishable port number are vulnerable to port-based server censorship. Among the evaluated schemes, only DNSCrypt, DoH, and DoTor employ port numbers that are shared other widely used protocols. Furthermore, it is evident that schemes susceptible to port-based censorship are also at risk of censorship based on IP address, as their name servers can be identified by Internet scanning techniques. Given that censoring organizations (agents) employ various strategies to restrict access to the Tor network~\cite{afroz2021timeline}, they can similarly block or degrade DoTor traffic along with other Tor-related communications. Additionally, DNSCrypt traffic has distinguishable characteristics~\cite{dnscryptIsolation}, making it easily recognizable and, therefore, susceptible to censorship. Overall, among the evaluated schemes, DoH appears to have the highest resilience against traffic isolation and server censorship.

\textbf{Availability:} Regarding the availability of recursive resolvers, only Stage~1 schemes that work on top of TCP or implement some means of source IP address verification before resolving a DNS query are \AResDoS. Additionally, such schemes also resist DNS-based reflection amplification attacks by verifying the source IP address. However, none of the evaluated Stage~2 schemes protects ANSes against DoS attacks. One possible explanation for the lack of DoS resistance schemes in Stage~2 is the limited number of in Stage~2, and ANSes may be reluctant to deploy stateful protocols that require maintaining the communication states and consequently result in significant per-query overhead compared to UDP-based schemes.

\textbf{Traffic distinguishability:} The distinguishability of DNS resolution traffic from other network traffic can harm the availability and privacy of DNS resolution. Two secure DNS alternatives satisfy some degree of \PConceal{} (\hyperlink{P2}{P2}), namely DoH and DoTor. However, it was shown that DoH traffic can be distinguished from web traffic~\cite{vekshin2020doh}; thus, DoH received partial credit for \PConceal{}. Therefore, only DoTor fully renders DNS traffic indistinguishable, and thereby mitigates attacks on isolated traffic of a DNS scheme, such as downgrade attacks or exclusive censorship of DNS traffic.

\textbf{Summary:} According to Table~\ref{tab:comptable}, Stage~1 protocols primarily provide confidentiality (\hyperlink{P1}{P1}), message authentication (\hyperlink{S1}{S1}), anti-replay mechanism (\hyperlink{S2}{S2}), resilience to payload-based censorship (\hyperlink{A1}{A1}), and resilience to resolver DoS (\hyperlink{A3}{A3}) in that stage. However, Stage~1 schemes often do not conceal the nature of DNS messages (\hyperlink{P2}{P2}) and most are susceptible to DNS server censorship (\hyperlink{A2}{A2}).

Regarding Stage~2 schemes, we observe that the number of Stage~2 schemes is limited, and the evaluated schemes provide a different set of properties compared to one another. Moreover, none of the schemes provides resilience to ANS DoS (\hyperlink{A4}{A4}); however, both of the evaluated schemes mitigate cache poisoning attacks by preventing false response injection (\hyperlink{S3}{S3}) and neither of the two hides clients IP addresses completely in Stage~2 (\hyperlink{P5}{P5}).

\section{Discussion}
In this section, based on our survey (Section~\ref{sec:alternatives}) and comparative evaluation, we discuss several high-level insights regarding the name resolution process. One general observation is that the schemes designed to work in Stage~2 are limited in that they provide only a handful of the defined security benefits. On the other hand, Stage~1 schemes are diverse, and provide various combinations of properties but none provides all.

\textbf{Slow migration away from Vanilla DNS.} Although numerous secure DNS schemes have been proposed over time, Vanilla DNS remains dominant for name resolution in practice, particularly in Stage~2~\cite{chung2017longitudinal, IanaDNS}. A subset of the secure DNS schemes have security weaknesses and deployability obstacles that impede their large-scale adoption. Therefore, designing new secure DNS schemes that are more secure and deployable appears to be necessary for widespread adoption. To enhance the adoption rate of secure DNS schemes, it appears crucial to incentivize and educate clients and administrators of recursive resolvers and ANSes on their benefits and implementation strategies. Furthermore, setting secure DNS schemes as the default configuration in client-side software (\eg web browsers) may be another effective approach to progress widespread adoption. However, for both current and future schemes, to prevent the centralized collection of client data in Stage~1, client software must, by default, use multiple resolvers from different organizations and distribute queries accordingly.

\textbf{The privacy benefits of TLS-based schemes are questionable.} TLS-based schemes (\ie DoT, DoH, DoQ, DoDTLS) rely on the evolved TLS and web PKI~\cite{madwebpaper}, but mainly provide security; in the DNS context, they often fail to satisfy privacy properties. For \hyperlink{P1}{P1}, encryption in TLS-based schemes must be enhanced with defense mechanisms (\eg strong padding) against traffic analysis attacks that reveal or allow identification of web pages visited by users. Additionally, most TLS-based schemes require means to conceal DNS message nature in Stage~1 (\hyperlink{P2}{P2}) and hide the IP addresses of clients from the recursive resolvers (\hyperlink{P3}{P3}). Aside from lack of privacy properties, the TLS-based schemes are primarily proposed to secure Stage~1, and Stage~2 remains unsecured. Taking into account these points, introducing the TLS-based schemes as secure and privacy-preserving schemes to clients risks misleading clients about name resolution security and privacy. For example, web browsers, such as Google Chrome or Microsoft Edge, introduce DoH as \emph{secure DNS}, but this only secures Stage~1 and does not provide any of our properties in Stage~2.

\textbf{Stronger privacy in other alternatives.} Regarding Stage~1 schemes, schemes that transmit DNS messages over the Tor network (\eg DoTor) provides more privacy benefits than others. DoTor is the only scheme that offers robust privacy against intermediate entities in Stage~1, as it provides strong confidentiality and conceals the nature of DNS traffic; thus, it prevents intermediate entities (including ISP) from detecting name resolution. Additionally, DoTor~\cite{dotor} hides the IP address of clients from recursive resolvers. To take advantage of the benefits of Tor and preserve security and privacy of DNS messages in Tor exit relays, other secure DNS schemes in Stage~1 (\eg DoT or DoH) could be combined with Tor when a regular (non-onion) recursive resolver is queried (\eg DoH-over-Tor~\cite{dotor}). However, using Tor results in significant latency, bandwidth increase, and computational overhead, which must be considered before integrating Tor with a secure DNS scheme. Aside from integrating DNS schemes with Tor, combining Stage~1 schemes with ODNS-like relays (\eg Oblivious DoH (ODoH)~\cite{rfc9230}) enhances their privacy by hiding client IP addresses from recursive resolvers. Among the Stage~2 schemes, DNSCurve illustrates stronger privacy properties mainly by utilizing authenticated encryption.

\textbf{ISP correlation.} ISPs can correlate encrypted DNS traffic and subsequent traffic originating from a client, and thus infer queried domain names from encrypted DNS messages. ISP correlation-based information gathering is enabled by two factors: the distinguishability of encrypted name resolution and domain name leaks from other protocols. This suggests that, a secure DNS scheme that renders its traffic indistinguishable (\ie provides \hyperlink{P2}{P2}) from other traffic can effectively mitigate these correlations. Mitigating domain name leaks in other protocols (\eg using ESNI or ECH~\cite{ESNIDraft}) can also alleviate such correlation-based domain name leaks.

\textbf{Refining Properties.} In this type of evaluation schemes as presented in Table~\ref{tab:comptable}, there is always room for the properties to be refined and evolved. Therefore, the future researchers can take into consideration more parameters in addition to the current definitions of properties. For example, our availability properties in Section~\ref{sec:properties} do not reflect the amplification factor of the schemes or protocol-specific computational resource exhaustion attacks (\eg KeyTrap attack~\cite{heftrig2024harder}).

In evaluating availability properties, we considered source address validation as a means in mitigating reflection amplification and DNS flooding attacks. Source address validation can be applied at the transport layer (\eg using TCP) or at the application layer by introducing an additional round-trip or maintaining state from previous interactions. However, these approaches inevitably incur memory, computational, or network overhead. To preserve the efficiency of single round-trip performance, some schemes defer source address validation until specific conditions are met, such as when the response size exceeds the query size or a predefined threshold. While this strategy provides an efficient defense against amplification attacks, these schemes remain vulnerable to DNS flooding attacks, as validation is not enforced from the beginning.

\textbf{Deployability issues.} Due to the two-staged structure of DNS resolution and the difference in required properties of each stage, proposing a comprehensive and efficient DNS scheme that provides security, privacy, and availability properties in both stages appears to be challenging. In addition, complex DNS alternatives that require significant changes or introduce considerable overhead are unlikely to achieve widespread adoption. Moreover, ANSes in Stage~2 have been reluctant to adopt schemes that are complex or add considerable time or computation overhead. Therefore, proposing a secure DNS scheme that meets the required properties of Stage~2 while overcoming deployability barriers to achieve large-scale adoption remains an open challenge.

\textbf{Combining schemes.} Since designing a single secure DNS scheme for both stages potentially results in complexity and significant changes to the name resolution process, combining schemes is a potential solution. There are two main approaches for combining secure DNS schemes: combining schemes within the same stage and combining schemes in different stages. Integrating schemes in the same stage augments the properties offered by the combined mechanisms, although it may also introduce efficiency and cost overheads. On the other hand, combining schemes in different stages secures both stages (the entire DNS resolution path) based on the selected schemes. For example, by combining DoT with DoTor~\cite{dotor}, or DoH with ODNS-like relays~\cite{rfc9230} in Stage~1, the resultant scheme increases the provided properties only in Stage~1. On the other hand, by integrating the TLS-based schemes in Stage~1 with DNSSEC in Stage~2, each scheme provides benefits in one stage, and by complementing each other the entire DNS resolution path is enhanced.

When combining schemes, the resulting overhead must not outweigh the augmented properties. For instance, while integrating Stage~1 schemes with Tor may enhance privacy, Tor introduces significant performance and computational overhead, making it unsuitable for latency-sensitive or resource-constrained scenarios.

\textbf{Centralization.} It is widely accepted that relying on a single resolver is harmful to clients' privacy in Stage~1~\cite{hounsel2020d, schmitt2019oblivious}. Accumulated name resolution history of clients can be used by resolution services for client identification and re-identification, and other goals, such as financial gains. The various types of stakeholders in a DNS resolution process, such as clients, network administrators, and organizations maintaining resolution services or ANSes may prioritize their objectives differently. However, prioritizing client privacy and security, a name resolution must use mechanisms that avoid centralization. For example, DoTor~\cite{dotor} employs Tor relays, and ODoH~\cite{rfc9230} or DNSCrypt~\cite{DNSCryptV3} use specific relays to conceal the IP address of clients from recursive resolvers. Beyond relay-based solutions, Hoang~\etal~\cite{hoang2020k} proposed the K-resolver to distribute DNS queries among different DoH resolvers, which mitigates the privacy risks of a single centralized resolver.

\textbf{Bypassing censorship.} Internet censoring entities often leverage DNS messages or resolvers to block access to specific services~\cite{dnsmanipulation,aryan2013internet}. Therefore, using non-censoring recursive resolvers, residing out of the authority of censoring agents, can help circumvent DNS-based censorship. The majority of schemes proposed for Stage~1 offer some degree of resilience against payload-based censorship. However, all these schemes remain vulnerable to server-based censorship, wherein censoring authorities can effectively block the entire scheme by blocking access to the associated servers.

\section{Concluding Remarks}
The evaluation framework presented in this paper offers a comprehensive overview of the current landscape of secure DNS schemes in both stages of DNS resolution. It reveals overarching patterns and trends that remain unrevealed when the schemes are examined individually. A notable pattern observed is that none of the existing schemes provides end-to-end security across the entire DNS resolution path. Given that the DNS resolution process typically operates in two distinct stages, a practical and effective approach would be to secure each stage individually by deploying an appropriate secure DNS scheme at each stage.

Regarding Stage~1 schemes, the TLS-basd secure DNS schemes in Stage~1 are promoted by both recursive resolvers and client-side software, and the majority effectively ensure the security and confidentiality of DNS messages. However, the remaining unsatisfied properties at this stage are primarily related to protecting privacy of clients against resolvers and concealing the nature of DNS messages to prevent the identification and isolation of DNS traffic, which is essential for resisting traffic analysis and network-based censorship attacks.

On the other hand, although recent literature demonstrates the feasibility of cache poisoning by new techniques, there is no widely adopted DNS scheme in Stage~2. Also, the employed schemes in Stage~1 were not suitable and accepted to be used in Stage~2. All schemes evaluated in Stage~2 effectively mitigate cache poisoning attacks. However, their reliance on UDP necessitates mechanisms for source verification and protection against DoS attacks. Furthermore, as these schemes increasingly aim to provide real-time security and privacy guarantees, an additional security and deployability requirement emerges, which is avoiding the duplication of long-term secret.

Overall, this survey has demonstrated that each stage of the DNS resolution process presents distinct security, privacy, and availability requirements, and we believe it will be instrumental in paving the path that shapes future development of secure DNS standards and practices.

\balance
\bibliographystyle{IEEEtran}
\bibliography{Bibliography}

\end{document}